\documentclass[reqno,12pt,a4paper]{article}
\usepackage{amssymb,amsmath}
\usepackage{color,cite}
\usepackage{accents}
\usepackage[title]{appendix}
\usepackage{texdraw}
\usepackage{eucal}
\usepackage{epic,epsfig}
\usepackage{graphicx}
\usepackage[all]{xy}
\usepackage{tikz}
\usepackage{mathtools}
\usepackage{mdframed}
\newtheorem{theorem}{Theorem}[section]
\newtheorem{lemma}[theorem]{Lemma}

\numberwithin{equation}{section}
\newcommand{\bse}{\begin{subequations}}
\newcommand{\ese}{\end{subequations}}
\newcommand{\wt}{\widetilde}
\newcommand{\wh}{\widehat}
\newcommand{\ut}{\undertilde}

\newcommand{\Cal}{\mathcal}

\newcommand{\bL}{\boldsymbol{L}}
\newcommand{\bM}{\boldsymbol{M}}
\newcommand{\bsu}{\boldsymbol{u}}
\newcommand{\bsU}{\boldsymbol{U}}

\newcommand{\be}{\begin{equation}}
\newcommand{\ee}{\end{equation}}
\newcommand{\bea}{\begin{eqnarray}}
\newcommand{\eea}{\end{eqnarray}}
\newcommand{\nn}{\nonumber}

\newcommand{\PDel}{{P$\Delta$E}}

\newcommand{\oa}{\omega}

\def \h#1{\widehat{#1}}
\def \t#1{\widetilde{#1}}
\def \th#1{\widehat{\widetilde{#1}}}
\def \b#1{\overline{#1}}
\def \c#1{\accentset{\circ}{#1}}
\def \d#1{\accentset{\bullet}{#1}}
\def \dc#1{\underaccent{\circ}{#1}}
\def \td#1{\accentset{\bullet}{\widetilde{#1}}}
\def \hd#1{\accentset{\bullet}{\widehat{#1}}}
\def \thd#1{\accentset{\bullet}{\widehat{\widetilde{#1}}}}

\def \ddc#1{\accentset{\bullet}{\underaccent{\circ}{#1}}}

\def \tdc#1{\widetilde{\underaccent{\circ}{#1}}}
\def \hdc#1{\widehat{\underaccent{\circ}{#1}}}
\def \tddc#1{\underaccent{\circ}{\accentset{\bullet}{\widetilde{#1}}}}
\def \hddc#1{\underaccent{\circ}{\accentset{\bullet}{\widehat{#1}}}}

\def \xht#1{\widehat{\widetilde{#1}}}
\def \xb#1{\overline{#1}}
\def \xbt#1{\overline{\widetilde{#1}}}
\def \xbh#1{\overline{\widehat{#1}}}
\def \xbht#1{\overline{\widehat{\widetilde{#1}}}}

\def\hypohat#1#2{\vrule depth #1 pt width 0pt{\smash{{\mathop{#2}
 \limits_{\displaystyle\widehat{}}}}}}

\newcommand{\uh}[1]{\hypohat 0 #1}
\newcommand{\PDE}{P$\Delta$E }

\title{Discrete Boussinesq-type equations}

\author{Jarmo Hietarinta,$^1$\footnote{E-mail: jarmo.hietarinta@utu.fi}
  ~and~ Da-jun Zhang$^2$\footnote{E-mail: djzhang@staff.shu.edu.cn }
  \\
  {\small\it $^1$Department of Physics and Astronomy,
    University of Turku, FIN-20014 Turku, Finland} \\
  {\small\it $^2$Department of Mathematics, Shanghai University,
    Shanghai 200444, P.R. China}}
\date{\today}

\begin{document}

\maketitle

\begin{abstract}
  We present a comprehensive review of the discrete Boussinesq
  equations based on their three-component forms on an elementary
  quadrilateral. These equations were originally found by Nijhoff et
  al using the direct linearization method and later generalized by
  Hietarinta using a search method based on multidimensional
  consistency.  We derive from these three-component equations their
  two- and one-component variants. From the one-component form we
  derive two different semi-continuous limits as well as their fully
  continuous limits, which turn out to be PDE's for the regular, modified and
  Schwarzian Boussinesq equations. Several kinds of Lax pairs are also
  provided.  Finally we give their Hirota bilinear forms and
  multi-soliton solutions in terms of Casoratians.


\end{abstract}

\tableofcontents

\section{Introduction}
Among the $1+1$ dimensional soliton equations there are evolution
equations, such as the Korteweg -- de Vries (KdV) equation, in which
time derivatives appear in first order, but there are also important
equations with higher order time derivatives, such as the Boussinesq
(BSQ) equation. An essential difference between these equations is in
the initial data required: For KdV it would be enough to give, e.g.,
$u(x,t=0)$, while for the second order BSQ equation we would need
$u(x,t=0)$ and $\partial_x u(x,t=0)$, or something similar.

The difference between the first and second order time evolution is
reflected also in the integrable discretizations of these equations.
For first order equations a well defined evolution is obtained from a
staircase-like initial data together with an equation defined on the
elementary square of the lattice. For higher order time evolutions one
would then need either initial data on a number of parallel staircases
with an equation on a larger stencil or alternatively,
multi-component initial data with a larger number of equations on the
small stencil.

The recent rapid advances in the study of integrable partial
difference equations (\PDel) are to a large extent due to the
efficient use of the particular integrability property of {\em
  multidimensional consistency} (MDC), which is related to the
existence of hierarchies in the continuous case \cite{HJN}.  In its
simplest form it involves dimensions 2 and 3 and is called
Consistency-Around-a-Cube (CAC).  The MDC property was discussed
already in \cite{NW-GLA-2001,N-PLA-2002,BS-IMRN-2002} but in full
force it was applied in \cite{ABS-CMP-2003} (with some further
technical assumptions), and this provided a classification of first
order equations defined on an elementary lattice square of the
Cartesian 2D lattice, the so called Adler-Bobenko-Suris (ABS) list.
The requirement of MDC can also be applied on multicomponent equations
on the elementary plaquette. A partial classification of
three-component equations was done in \cite{H-JPA-2011} on the basis
of CAC and most of the results turned out to be discrete versions BSQ
equations (DBSQ).

Multicomponent equations were also studied from the perspective of
direct linearization approach (DLA) and several equations were found
\cite{NQC-PLA-1983,QNCL,NPCQ-IP-1992,W-PhD-2001,ZZN-SAM-2012,FuN-PRSA-2017}.
In addition to the CAC and DLA approaches DBSQ equations have been
derived also by applying a three-reduction on the three-term Hirota-Miwa
equation \cite{djm388,MK}, or on the four-term Miwa equation
\cite{HZ-JDEA-2013}. Still further results have been obtained using
the Cauchy matrix approach \cite{FZZ-JNMP-2012} or graded Lax pairs
\cite{FX-13}.

In this review we will discuss in detail the multi-component DBSQ
equations. In Section \ref{sec-3} we compare the various three-component
forms that have appeared in the literature and their connections by
gauge transformations or by M\"obius transformations.  Their
symmetries are also briefly discussed. In fact, all the DBSQ-type
equations found in \cite{H-JPA-2011} can be viewed as extensions of
some known lattice equations found in 1990's. In Section \ref{S:1comp}
we discuss how the dynamics of the first order three-component
equations can be represented by two- or one-component forms on a
larger stencil. We also discuss the continuum limits of the
one-component forms in Section \ref{sec-3.3} and show that the limits
really are BSQ-like equations.

In Section \ref{S:lax} we present the Lax pairs of the various
discrete forms.  As for the solutions of DBSQ-type equations, besides the
results from direct linearisation and Cauchy matrix approach (see
\cite{NPCQ-IP-1992,W-PhD-2001,TN-GMJ-2005,ZZN-SAM-2012,FZZ-JNMP-2012}),
equation B2 has been bilinearized in \cite{HZ-JMP-2010} and solutions
were given in terms of Casoratians; in Section \ref{sec-4} we
investigate bilinear forms for A2 and C3 equations.

\section{DBSQ-type equations}\label{sec-3}
\subsection{Basic concepts and definitions}
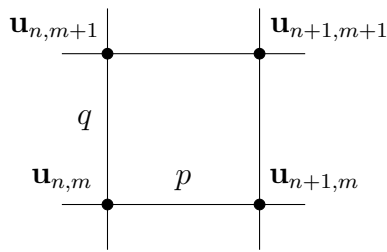
\begin{figure}[h]
\centering
\begin{tikzpicture}[scale=2.0]
 \draw[thin] (-0.3,0) -- (1.3,0);
 \draw[thin] (-0.3,1) -- (1.3,1);
 \draw[thin] (0,-0.3) -- (0,1.3);
 \draw[thin] (1,-0.3) -- (1,1.3);
\filldraw [black] (1,1) circle (1pt);
\filldraw [black] (0,0) circle (1pt);
\filldraw [black] (1,0) circle (1pt);
\filldraw [black] (0,1) circle (1pt);
\node at (-0.3,0.16) {${\mathbf u}_{n,m}$};
\node at (1.36,0.16) {${\mathbf u}_{n+1,m}$};
\node at (0.5,0.16) {$p$};
\node at (-0.35,1.16) {${\mathbf u}_{n,m+1}$};
\node at (-0.15,0.56) {$q$};
\node at (1.46,1.16) {${\mathbf u}_{n+1,m+1}$};
\end{tikzpicture}
\caption{Elementary  quadrilateral  of   the  lattice,  with  possibly
  multi-component corner variables ${\mathbf u}$. The parameters $p,q$
  characterize the distance between lattice points. \label{F:2}}
\end{figure}

The discrete equations that we discuss here are all defined on the
Cartesian $\mathbb Z\times\mathbb Z$ lattice. Most of the time the
equations are defined on a single quadrilateral but larger stencils
are sometimes needed. The independent variables live on the
vertices of the lattice and are therefore labeled by the vertex
coordinates, see Figure \ref{F:2}.

Sometimes the equations have a simpler look if we replace the
subscript with a tilde or a hat, or use some other simplified notation,
for example
\begin{equation}
   {\mathbf u}={\mathbf u}_{n,m}\!={\mathbf u}_{0,0},\
  \t{{\mathbf u}}={\mathbf u}_{n+1,m}\!={\mathbf u}_{1,0},\
  \h{{\mathbf u}}={\mathbf u}_{n,m+1}\!={\mathbf u}_{0,1},\
  \th{{\mathbf u}}={\mathbf u}_{n+1,m+1}\!={\mathbf u}_{1,1}.
\label{th}
\end{equation}

The equation(s) on the quadrilateral are given by ${\mathbf Q}
({\mathbf u}, \t{{\mathbf u}},\h{{\mathbf u}},\th{{\mathbf u}})=0$,
where ${\mathbf Q}$ are affine multilinear polynomials, and one may
then ask whether the system of equations is integrable according to
some definition. We will use the MDC criterion which means that the
equation defined on the 2D-lattice can be extended consistently into
higher dimensions.

As an example consider the simple case of the 1-component lattice
potential KdV equation (H1 in the ABS list):
\[
Q(u,\t u,\h u,\xht u;p,q):=(u-\th u)(\h u-\t u)-p^2+q^2=0.
\]
Since $u$ is defined on the 2D-lattice it naturally depends only on the
coordinates $n,m$, but in order to extend the equations into 3D setting,
which is the simplest requirement of MDC, we introduce a third
variable $k$ and denote $u_{n,m,k+1}=\b u$; the associated lattice
parameter is $r$. For each of the six sides of the cube we have an
equation: \bse\label{eq:conseqs}\begin{align} \text{bottom:}\quad &
  Q(u,\t u,\h u,\xht u;p,q)=0,& \text{top:}\quad &Q(\xb u,\xbt u,\xbh
  u,\xbht u;p,q)=0,&\\ \text{back:}\quad & Q(u,\h u,\xb u,\xbh
  u;\alpha,\beta)=0,& \text{front:} \quad &Q(\t u,\xht u,\xbt u,\xbht
  u;\alpha,\beta)=0,&\\ \text{left:}\quad & Q(u,\xb u,\t u,\xbt
  u;\gamma,\delta)=0,& \text{right:} \quad &Q(\h u,\xbh u,\xht u,\xbht
  u;\gamma,\delta)=0.&
\end{align}
\ese (Note that we used arbitrary parameters in this example and hope
to determine them by consistency.) In order to study CAC we choose
$u,\t u,\h u,\xb u$ as initial values and then from the equations on
the LHS we can compute $\xht u,\,\xbh u,\,\xbt u$. When these are used
on the RHS we have three ways to compute $\xbht u$ but they must all
yield the same value.  In this case we find the condition
\[
(\alpha^2-\beta^2)+(\gamma^2-\delta^2)+(p^2-q^2)=0.
\]
Since back and front equations do not depend on $p$ we find
$(\alpha^2-\beta^2)=(q^2-r^2)$ and $(\gamma^2-\delta^2)=r^2-p^2$ for
some $r$ and then all RHS equations yield the very symmetric form
\[
\xbht u=\frac{\h u\t u(q^2-p^2)+\t u\b u(p^2-r^2)+\b u\h u(r^2-q^2)}
      {\b u(p^2-q^2)+\h u(r^2-p^2)+\t u(q^2-r^2)}.
\]
Note that the RHS of this expression does not depend on $u$, this is
called the tetrahedron property.

The application of MDC on one-component equations resulted in the
ABS-list \cite{ABS-CMP-2003}. For multi-component equations there are
no equally comprehensive classifications. It should also be noted that
passing the CAC test is necessary but not sufficient \cite{JH2019} for
integrability.

\subsection{Hietarinta's list of equations}
In \cite{H-JPA-2011} Hietarinta made a partial classification of
Boussinesq (BSQ) type lattice equations using CAC. Since BSQ equations are of
second order in time, their discrete analogues are either
multi-component on a quadrilateral, or defined on a larger stencil. In
\cite{H-JPA-2011} three-component approach was used and some equations
were defined on the links and one equation on the full quadrilateral.
Using CAC, the following three-component DBSQ-type
equation were found: \bse\label{B2-H}
\begin{flalign}
\hbox{B2:}\hskip 3cm & \wt{y} = x\wt{x}-z,& \label{B2-H-a}\\
& \wh{y} = x\wh{x}-z, \label{B2-H-b}\\
& y = x\wh{\wt{x}}-\wh{\wt{z}} +b_0(\wh{\wt{x}}-x)+b_1+\frac{P-Q}{\wt{x}-\wh{x}}, \label{B2-H-c}
\end{flalign}
\ese
\bse\label{A2-H}
\begin{flalign}
\hbox{A2:}\hskip 3cm & \wt{y} = z\wt{x}-x,& \label{A2-H-a}\\
& \wh{y} = z\wh{x}-x, \label{A2-H-b} \\
& y =x\wh{\wt{z}}-b_0
x+\frac{P \wt{x}-Q \wh{x}}{\wh{z}-\wt{z}},~~~~~~~~~~~~~~~~\label{A2-H-c}
\end{flalign}
\ese
\begin{subequations}\label{C3-H}
\begin{flalign}
\hbox{C3:}\hskip 3cm & \t y \,z=\t x - x,&\label{C3-H-a}\\
 &\h y \,z=\h x - x,\label{C3-H-b}\\
 &\th{z}\,y= b_0\,x + b_1
        + z \frac{P\,\t y\,\h z-Q\,\h y\,\t z}{\t z-\h z},~~~~~~~~\label{C3-H-c}
\end{flalign}
\end{subequations}
and
\begin{subequations} \label{C4-H}
\begin{flalign}
\hbox{C4:}\hskip 3cm & \t y \,z =\t x - x,&\\
    &     \h y \,z=\h x - x,\\
    &     \th{z}\,y=  x\,\th x + b_2+ z\,\frac{P\,\t y\,\h z-Q\,\h y\,\t z}{\t z-\h z}.~~~~~~~~~
\end{flalign}
\end{subequations}
Here the parameters $P$ and $Q$ are related to lattice spacing
parameters $p,q$ in $n$ and $m$-directions, respectively.

The convention for naming the variables was designed for MDC and for
analyzing the evolution (Section \ref{S:evo}): The quasilinear equations
are defined on the edges of the quadrilateral; ``a'' equations depend
always on $x$, $\t x$, $z$, and $\t y$, and the ``b'' equations on
$x$, $\h x$, $z$, and $\h y$, i.e., the dependence is always the same,
only the algebra is different. Also, we list the equations in the
order B2, A2, C3,4 because, as we will see later, they correspond to
regular, modified and Schwarzian BSQ equations, respectively.

The coupling constants $b_i$ are arbitrary and generalize some
previous results. Note however, that $b_1$ in B2 can be removed
with the transformation
\begin{equation}
(x,y,z) \mapsto (x, y-\frac{b_1}{3}(n+m-1), z+\frac{b_1}{3}(n+m)),
\label{B2-tr-b1}
\end{equation}
and $b_0$ in A2 can be removed using
\begin{equation}
(x,y,z) \mapsto (x, y+\frac{b_0 x}{3}(n+m), z+\frac{b_0}{3}(n+m+1)).
\label{tr-A2-b0}
\end{equation}
In the following we do not keep these removable parameters.

In addition some two-component forms were found in \cite{H-JPA-2011}:
\begin{subequations} \label{C2-1-H}
\begin{flalign}
\hbox{C2-1:}\hskip 3cm & \th x =\frac{\h x \t z-\t x \h z}{\t z-\h z},&
\label{C2-1-H-a}\\
    &     \th{z}= -{b_0}z\,\th x + z\,\frac{P\,\h z-Q\,\t z}{\t z-\h z},
\label{C2-1-H-b}
\end{flalign}
\end{subequations}
and
\begin{subequations} \label{C2-2-H}
\begin{flalign}
\hbox{C2-2:}\hskip 3cm & \th x =\frac{\h x \t z-\t x \h z}{\t z-\h z},&
\label{C2-2-H-a}\\
    &    x \th{z}= -b_0 z + z\,\frac{P\,\t x\,\h z-Q\,\h x\,\t z}{\t z-\h z}.
\label{C2-2-H-b}
\end{flalign}
\end{subequations}
These are actually discrete versions of KdV so we will not discuss
them further here.

\subsection{Relations between the C-equations\label{SS:C34}}
Let us first note that lattice equations are classified only up to
local rational-linear (i.e., M\"obius) transformations, and that equations
related by them are considered same. However, for some purposes a
particular form may be better in practice.

Note that for the C-equations one can derive relation \eqref{C2-1-H-a}
by eliminating $y$. After the transformations discussed below it is
often useful to use this relation when comparing results.

First note that  if $b_0 \neq 0$ then by the transformation
\begin{equation}\label{tr-C3-b0b1}
  x \rightarrow x- \frac{b_1}{b_0},
\end{equation}
one can remove from the C3 equation \eqref{C3-H-c} the parameter
$b_1$ and then we can consider the following form:
\begin{subequations}\label{C3-H-1}
\begin{align}
\t y \,z &=\t x - x,\label{C3-H-1-a}\\
\h y \,z &=\h x - x,\label{C3-H-1-b}\\
\th{z}\,y &= b_0\,x
        + z \frac{P\,\t y\,\h z-Q\,\h y\,\t z}{\t z-\h z},\label{C3-H-1-c}
\end{align}
\end{subequations}
which we call C3$_{b_0}$.

Since the transformation \eqref{tr-C3-b0b1} fails when $b_0=0$, the
following equation
\begin{subequations}\label{C3-H-b1}
\begin{align}
\t y \,z &=\t x - x,\label{C3-H-b1-a}\\
\h y \,z &=\h x - x,\label{C3-H-b1-b}\\
\th{z}\,y &= b_1
        + z \frac{P\,\t y\,\h z-Q\,\h y\,\t z}{\t z-\h z}\label{C3-H-b1-c}
\end{align}
is not a trivial subcase of C3 equation \eqref{C3-H}. Since \eqref{C3-H-b1-c} does not contain $x$, we get a two-component form after eliminating  $x$ from
(\ref{C3-H-b1-a}, \ref{C3-H-b1-b}) and their shifts, this results in
\begin{equation}\label{C3-H-b1-d}
\th{y}=-z\frac{\t y-\h y}{\t z-\h z}.
\end{equation}
\end{subequations}
Thus (\ref{C3-H-b1-c},\ref{C3-H-b1-d}) is a two component form, let us
call it C3$_{b_1}$.

We will next show that C4 can be obtained from C3 by a M\"obius
transformation. As the first step we note that from equations
(\ref{C3-H-a},\ref{C3-H-b}) and their shifts one can derive
\begin{equation}\label{C3-H-link2-2-1}
  x = \th x + z\,\frac{\t y \h z - \h y \t z}{\t z - \h z}.
\end{equation}
Using it to replace $\tfrac12b_0x$ in \eqref{C3-H-1-c}
we get the following alternative form for C3$_{b_0}$:
\begin{subequations} \label{C3-H-3}
\begin{align}
  \t y \,z &=\t x - x,\label{C3-H-3-a}\\ \h y \,z &=\h x -
  x,\label{C3-H-3-b}\\ \th{z}\, y &= z\,\frac{(P-c_2)\,\t y\,\h
    z-(Q-c_2)\,\h y\,\t z}{\t z-\h z}+ c_2 (x+\th x), \quad
  (c_2=\frac{b_0}{2}).
  \label{C3-H-3-c}
\end{align}
\end{subequations}

Now, inserting the (mixed) M\"obius transformation\cite{ZZN-SAM-2012}
\begin{equation}
x=\frac{x_1-c_2}{2c_2(x_1+c_2)}, \quad
y=\frac{y_1}{x_1+c_2}, \quad
z=\frac{z_1}{x_1+c_2},
\label{trans-C3-C4}
\end{equation}
into \eqref{C3-H-3} we get
\begin{subequations} \label{C4-H-1}
\begin{align}
 \t y_1 \,z_1 &=\t x_1 - x_1,\\
 \h y_1 \,z_1 &=\h x_1 - x_1,\\
 \th{z}_1\,y_1
&= z_1\,\frac{(P-c_2)\,\t y_1\,\h z_1-(Q-c_2)\,\h y_1\,\t z_1}
             {\t z_1-\h z_1}
 + x_1\,\th x_1 - c_2^2,
\end{align}
which is C4$_{b_2}$ equation \eqref{C4-H}, after redefining
\end{subequations}
\begin{equation}
P\to P-c_2, \quad Q\to Q-c_2, \quad b_2=-c_2^2.
\end{equation}

The above transformation fails if $b_0=0$ in C3 i.e., if $b_2=0$ in
C4, but that special case can be obtained from C3${}_{b_1=1}$ by
the following transformation:
\begin{equation}\label{C4-into-3-b'}
x=-1/x_1,\quad y=y_1/x_1,\quad z=z_1/x_1.
\end{equation}
But since C3${}_{b_1=1}$ depends only on 2 variables, the
transformation \eqref{C4-into-3-b'} in fact eliminates the $x$ variable.

In summary, among the C-equations we only need to consider the
three-component equation C3${}_{b_0}$ \eqref{C3-H-1} for $b_0\neq
0$, and the two-component equations C3${}_{b_1}$
(\ref{C3-H-b1-c},\ref{C3-H-b1-d}) (for arbitrary $b_1$).

\subsection{Symmetries}

\subsubsection{$n\leftrightarrow m$ reflection symmetry}
As can be easily seen, all the equations are invariant under the
$n\leftrightarrow m$ reflection, i.e., $\t {}\, \leftrightarrow\, \h {}$ ,
accompanied by $P\leftrightarrow Q$ parameter change.

\subsubsection{Reversal symmetry\label{S:rev-sym}}
By reversal symmetry we mean symmetry under changing all tildes to
undertildes and hats to underhats. More precisely, the indices change
sign and then the generic point is renamed:
\[
x_{n+\nu,m+\mu} \mapsto  x_{-n-\nu,-m-\mu}=x_{n'-\nu,m'-\mu},
\]
after which we can drop the primes. This reversal is then with respect
to the lattice point $(n,m)$. In the notation where only shifts
relative to $(n,m)$ are indicated (such as $x_{0,1}$) we have $
x_{\nu,\mu} \mapsto x_{-\nu,-\mu},$ after which we usually shift the
whole equation.

\paragraph{B2:} If we apply this reversal to B2
equation \eqref{B2-H-a} we have
\[
 \wt{y} =
 x\wt{x}-z\quad \xmapsto{\phantom{m}\text{reversal}\phantom{m}} \quad \ut{y}
 = x\ut{x}-z\quad \xmapsto{\phantom{m}\text{shift}\phantom{m}} \quad
 y=\wt{x}x-\wt{z}.
\]
Thus we have reversal symmetry of \eqref{B2-H-a} (and \eqref{B2-H-b})
if we add the exchange $y \leftrightarrow z$. As for
\eqref{B2-H-c}, it turns out that we should also take
$b_0\leftrightarrow -b_0$ and $(P,Q) \leftrightarrow (-P,-Q)$
which is natural for a reversal of direction.  In summary, the B2
equations are reversal invariant if accompanied with
\begin{equation}\label{eq:rev-var-B}
(x,y,z,P,Q,b_0)\mapsto (x,z,y,-P,-Q,-b_0).
\end{equation}

\paragraph{C:}
Similarly for the C3 equations
we have reversal symmetry if we include variable changes
\begin{equation}\label{eq:rev-var-C}
(x,y,z)\mapsto (-x,z,y).
\end{equation}
For (\ref{C3-H-a}, \ref{C3-H-b})
this is manifest, and also for most terms in \eqref{C3-H-c} but some
terms need more computations. For example we have
\[
z \frac{P\,\t y\,\h z-Q\,\h y\,\t z}{\t z-\h z}
~ \mapsto ~
z \frac{P\,\ut y\,\uh{z}-Q\,\uh y\,\ut z}{\ut z-\uh z}
~ \mapsto ~
\h{\t z} \frac{P\,\h y\,\t{z}-Q\,\t y\,\h z}{\h z-\t z}
~ \mapsto ~
\h{\t y} \frac{P\,\h z\,\t{y}-Q\,\t z\,\h y}{\h y-\t y},
\]
and to finish the computation we should still show that
${z}/({\t z-\h z})={\h{\t y}}/({\h y-\t y})$
but this follows by taking suitable linear combination
of \eqref{C3-H-a}, \eqref{C3-H-b} and their shifts.

The only other special term in C3 is $b_0\,x$ in \eqref{C3-H-c}
which in this process changes to $-b_0\h{\t x}$. However, by taking
again a suitable combination of the (\ref{C3-H-a}, \ref{C3-H-b}) and
their shifts we can derive
\[
\h{\t x}=x-z\frac{\t y\h z-\h z\t y}{\t z-\h z},
\]
which combines with the $P,Q$ term. Thus if $b_0\neq0$, we have
reversal symmetry if we do the further parameter replacements
\begin{equation}\label{eq:C-rev-par}
(P,Q,b_0,b_1)\mapsto(P+b_0,Q+b_0,-b_0).
\end{equation}

The two-component equation C3$_{b_1}$ of
(\ref{C3-H-b1-c},\ref{C3-H-b1-d}) is also reversal symmetric.

\paragraph{A2:} The case of A2 is a bit more complicated:
From \eqref{A2-H-a} we have
\[
 \wt{y} = z\wt{x}-x\quad \mapsto \quad
 \ut{y} = z\ut{x}-x\quad \mapsto \quad
y=\wt{z}x-\wt{x}
\]
but the usual map \eqref{eq:rev-var-C} does not take this to the
original form. Dividing the last equation by $x\wt x$ yields
$y/(x\wt{x})=\wt{z}/\wt{x}-1/x$. Now it can be seen that we get the
original form if instead of \eqref{eq:rev-var-C} we have
\begin{equation}\label{eq:rev-var-A2}
(y,z,x)\mapsto (-z/x,-y/x,1/x).
\end{equation}
For the $P,Q$ term we only need to show that $\h{\t x}x/(\h y\t
x-\t y\h x)=1/(\h z-\t z)$, which follows from
(\ref{A2-H-a},\ref{A2-H-b}) and their shifts. The
form \eqref{eq:rev-var-A2} suggest that the transformation can be
simplified if we use another variable $w:=y/x$. Then we get the
alternate form
\bse\label{A2-Hw}
\begin{flalign}
\hbox{A2-alt:}\hspace{4cm} z-\wt{w}
=& \frac{x}{\wt{x}}, &\label{A2-Hw-a}\\
 z-\wh{w} =& \frac{x}{\wh{x}}, \label{A2-Hw-b} \\
(\wt{z}-\wh{z})(\wh{\wt{z}}-w)=&
P\,\frac{\t x}{x}-Q\,\frac{\h x}{x}.\label{A2-Hw-c}
\end{flalign}
\ese
These are reversal
invariant with
\begin{equation}\label{eq:rev-var-A2w}
(w,z,x)\mapsto (-z,-w,1/x).
\end{equation}
This is easy to show if one uses the formula
\[
\frac{\wh{\wt{x}}\,x}{\t x\,\h x}=\frac{\t w-\h w}{\t z-\h z}
\]
which follows from (\ref{A2-Hw-a},\ref{A2-Hw-b}).

\subsection{Initial values and evolution\label{S:evo}}
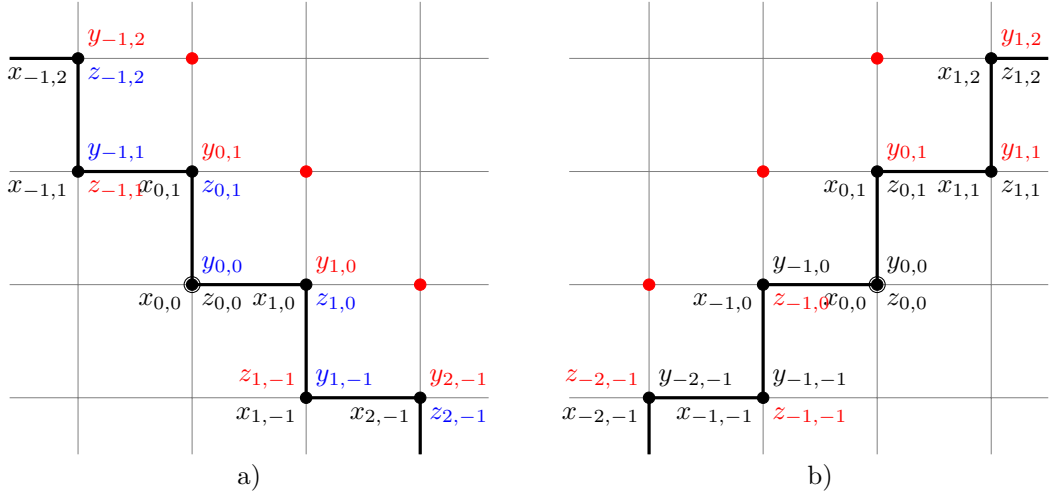
\begin{figure}
{\footnotesize
\centering
\begin{tikzpicture}[yscale=1.5,xscale=1.5]
\draw[step=1cm,gray,very thin] (-0.6,1.5) grid (3.6,5.5);
\draw[very thick] (-0.6,5) -- (0,5) -- (0,4) -- (1,4) -- (1,3) -- (2,3)
 -- (2,2) -- (3,2) -- (3,1.5);
\filldraw[red] (2,4) circle (0.05cm);
\filldraw[red] (3,3) circle (0.05cm);
\filldraw[red] (1,5) circle (0.05cm);
\draw[black] (1,3) circle (0.07cm);
\filldraw[black] (1,3) circle (0.05cm) node[below left] {$x_{0,0}$} ;
\draw[black] (1,3) node[below right] {$z_{0,0}$} ;
\draw[blue] (1,3) node[above right] {$y_{0,0}$} ;
\filldraw[black] (2,3) circle (0.05cm) node[below left] {$x_{1,0}$} ;
\draw[red] (2,3) node[above right] {$y_{1,0}$} ;
\draw[blue] (2,3) node[below right] {$z_{1,0}$} ;
\filldraw[black] (1,4) circle (0.05cm) node[below left] {$x_{0,1}$} ;
\draw[red] (1,4) node[above right] {$y_{0,1}$} ;
\draw[blue] (1,4) node[below right] {$z_{0,1}$} ;
\filldraw[black] (0,4) circle (0.05cm) node[below left] {$x_{-1,1}$} ;
\draw[red] (0,4) node[below right] {$z_{-1,1}$} ;
\draw[blue] (0,4) node[above right] {$y_{-1,1}$} ;
\filldraw[black] (0,5) circle (0.05cm) node[below left] {$x_{-1,2}$} ;
\draw[red] (0,5) node[above right] {$y_{-1,2}$} ;
\draw[blue] (0,5) node[below right] {$z_{-1,2}$} ;
\filldraw[black] (2,2) circle (0.05cm) node[below left] {$x_{1,-1}$} ;
\draw[red] (2,2) node[above left] {$z_{1,-1}$} ;
\draw[blue] (2,2) node[above right] {$y_{1,-1}$} ;
\filldraw[black] (3,2) circle (0.05cm) node[below left] {$x_{2,-1}$} ;
\draw[red] (3,2) node[above right] {$y_{2,-1}$} ;
\draw[blue] (3,2) node[below right] {$z_{2,-1}$} ;
\draw[black] (1.5,1.5) node[below] {a)} ;
\end{tikzpicture}
\hspace{0.5cm}
\begin{tikzpicture}[yscale=1.5,xscale=1.5]
\draw[step=1cm,gray,very thin] (-0.7,1.5) grid (3.5,5.5);
\draw[very thick] (3.5,5) -- (3,5) -- (3,4) -- (2,4) -- (2,3) -- (1,3)
 -- (1,2) -- (0,2) -- (0,1.5);
\filldraw[red] (1,4) circle (0.05cm);
\filldraw[red] (0,3) circle (0.05cm);
\filldraw[red] (2,5) circle (0.05cm);
\filldraw[black] (2,3) circle (0.05cm) node[below left] {$x_{0,0}$} ;
\draw[black] (2,3) node[below right] {$z_{0,0}$} ;
\draw[black] (2,3) node[above right] {$y_{0,0}$} ;
\filldraw[black] (1,3) circle (0.05cm) node[below left] {$x_{-1,0}$} ;
\draw[black] (1,3) node[above right] {$y_{-1,0}$} ;
\draw[red] (1,3) node[below right] {$z_{-1,0}$} ;
\filldraw[black] (2,4) circle (0.05cm) node[below left] {$x_{0,1}$} ;
\draw[red] (2,4) node[above right] {$y_{0,1}$} ;
\draw[black] (2,4) node[below right] {$z_{0,1}$} ;
\filldraw[black] (3,4) circle (0.05cm) node[below left] {$x_{1,1}$} ;
\draw[black] (3,4) node[below right] {$z_{1,1}$} ;
\draw[red] (3,4) node[above right] {$y_{1,1}$} ;
\filldraw[black] (3,5) circle (0.05cm) node[below left] {$x_{1,2}$} ;
\draw[red] (3,5) node[above right] {$y_{1,2}$} ;
\draw[black] (3,5) node[below right] {$z_{1,2}$} ;
\filldraw[black] (1,2) circle (0.05cm) node[below left] {$x_{-1,-1}$} ;
\draw[red] (1,2) node[below right] {$z_{-1,-1}$} ;
\draw[black] (1,2) node[above right] {$y_{-1,-1}$} ;
\filldraw[black] (0,2) circle (0.05cm) node[below left] {$x_{-2,-1}$} ;
\draw[black] (0,2) node[above right] {$y_{-2,-1}$} ;
\draw[red] (0,2) node[above left] {$z_{-2,-1}$} ;
\draw[black] (1.5,1.5) node[below] {b)} ;
\draw[black] (2,3) circle (0.07cm);
\end{tikzpicture}}
\caption{The initial values given on the staircase.  The black
  variables must be given in order to compute the red variables on the
  staircase using the quasilinear equations. In order to compute the
  values at red dots (one step in the evolution) it is also necessary
  to give the blue variables in case a).
\label{F1}}
\end{figure}

We will consider evolution starting from initial values given on a
staircase, on which the inside corner points are given by $n+m=0$ and
the outside corner points by $n+m=1$, see Figure \ref{F1} a). Another
possible initial staircase is given by $n-m=0$, $n-m=-1$, see Figure
\ref{F1} b). We have three sets of variables and several equations so
we must look carefully which kind of initial values are necessary and
make sense.

Let us first consider a staircase in the NW-SE direction,
Figure \ref{F1} a). We assume that $x$ is given on all points of the
staircase and to get started, also $z_{0,0}$. Then it is possible to
compute, step by step, the red $y$ and $z$ values using the
quasilinear equations, i.e., we can compute $y_{n,m}$ for $ n+m=1$ and
$z_{n,m}$ for $n+m=0,\,n\neq0$.  If the staircase is in the NE-SW
direction, as in Figure \ref{F1} b), it is necessary to give $z_{n,m}$
for $n-m=0,-1,\,n\ge0$ and $y_{n,m}$ for $n-m=0,-1,\,n\le0$.

After the above we still have three equations left for each square, by
which we should be able to compute values at points where $n+m=2$ (red
dots in Figure \ref{F1}). But before that can be done for the staircase a)
we need more initial values: In order to use the remaining quasilinear
equations we also need $z_{n,m},\,n+m=1$, and for the fully nonlinear
equation, $y_{n,m},\, n+m=0$. These additional necessary initial
values are given in blue in Figure \ref{F1}. For staircase b) all
necessary initial values were needed already for filling in the
staircase.

Since the systems are reversal symmetric the same initial values work
for evolution in the opposite direction.

\subsection{Connection with the direct-linearization results}\label{sec-3.1}
\subsubsection{The direct-linearization approach}
Many of the discrete Boussinesq equations discussed here were derived
earlier (see\cite{NPCQ-IP-1992,W-PhD-2001,TN-GMJ-2005,N-OUP-1999})
using the direct linearization scheme (DLA) of Capel, Nijhoff, Quispel
et al. The results of Hietarinta \cite{H-JPA-2011} provided
generalizations to the early results, but they were subsequently also
derived from the DLA point of view in \cite{ZZN-SAM-2012} by
generalizing the dispersion relation.

The DLA was first proposed by Fokas and Ablowitz \cite{FA-PRL-1981},
and soon was developed to the study of discrete integrable systems
\cite{NQC-PLA-1983,QNCL,NPCQ-IP-1992}.  In this approach, an infinite
matrix is introduced via a linear integral equation with certain plane
wave factors and discrete equations arise as closed forms of the shift
relations of the elements of the matrix.

For the DBSQ equations, one first introduces an integral equation for
infinite order column vector $\bsu(k)$:
\[\boldsymbol{u}(k)+\rho(k) \sum_{j=1}^2\int_{\Gamma_j} d\mu_j(l)\,
\bsu(l) \sigma(-\omega_j(l))=\rho(k) \boldsymbol{c}^{T}_k,\]
where $\Gamma_j$ and $d\mu_j(l)$ are contours and measures that need to be suitably chosen,
$\boldsymbol{c}_k$ is an infinite order constant column vector
$(\cdots, k^{-2}, k^{-1}, 1, k, k^2, \cdots)^T$,
plane wave factors are
\[\rho(k)=(p+k)^n(q+k)^m\rho^{(0)}(k),~~
\sigma(k')=(p-k')^{-n}(q-k')^{-m}\sigma^{(0)}(k'),\]
$\rho^{(0)}, \sigma^{(0)}$ are constants, $\omega_j(k)$
are defined through
\[p^3+\alpha_2 p^2-(k^3+\alpha_2 k^2)=(p-k)(p-\omega_1(k))(p-\omega_2(k)).\]
Then, introduce an $\infty\times \infty$ matrix $\bsU$ by
\[\bsU= \sum_{j=1}^2\int_{\Gamma_j} d\mu_j(l)\,
\sigma(-\omega_j(l)) \bsu(l)\,\boldsymbol{c}_{-\oa_j(l)}^T.\]
After that one can define scalar functions
\[S^{(i,j)}_{(a,b)}=\boldsymbol{e}^T(a+\Lambda)^{-1} \bsU (b-\Lambda)^{-1}\boldsymbol{e},
~~ i,j\in \mathbb{Z},~~ a,b\in \mathbb{C},\]
where $\boldsymbol{e}=(\cdots,0,0,1,0,0,\cdots)^T$ in which only the center element is $1$, and
$\Lambda=(\lambda_{i,j})_{\infty\times\infty}$ in which
$\lambda_{i,j}=\delta_{i,j-1}$.
More explicitly we have
\begin{align*}
& u^{(i,j)}= (-1)^j S^{(i,j)}_{(0,0)},~~ s_{a,b} = S^{(-1,-1)}_{(a,b)},~~
 v_a  = 1-S^{(-1,0)}_{(a,0)},~~w_b = 1+S^{(0,-1)}_{(0,b)}, \\
& s_a  = a+S^{(-1,1)}_{(a,0)},~~t_b =-b+S^{(1,-1)}_{(0,b)}, ~~
 r_a  = a^2-S^{(-1,2)}_{(a,0)},~~z_b = b^2+S^{(2,-1)}_{(0,b)}.
\end{align*}
DBSQ equations arise from closed forms of the shift relations of the above elements
(cf.\cite{ZZN-SAM-2012}).

Due to the different origin, the equations from DLA appear in a
different gauge and in this section we elaborate the connections.

\subsubsection{B2}\label{sec-3.1.1}

Let us focus on B2 equation \eqref{B2-H}, in which we may assume that
$b_1=0$, as mentioned before. If we use on \eqref{B2-H} the
transformation \cite{ZZN-SAM-2012} \bse
\label{tr-B2-H-ZZN}
\begin{align}
& x = u^{(0,0)}-x_{0}, \label{tr-B2-H-ZZN-x}\\
& y = u^{(0,1)}-x_{0}u^{(0,0)}+y_0,\\
& z = u^{(1,0)}-x_{0}u^{(0,0)}+z_0,
\end{align}
\ese
where
\bse
\label{0ss-B2}
\begin{align}
& x_0 = np+mq+c_1, \\
& y_0 = \frac{1}{2}(np+mq+c_1)^2-\frac{1}{2}(np^2+mq^2+c_2)-c_3,\\
& z_0 = \frac{1}{2}(np+mq+c_1)^2+\frac{1}{2}(np^2+mq^2+c_2)+c_3
\end{align}
\ese
and $c_j  (j=1,2,3)$ are constants, we obtain
\bse\label{B2-ZZN}
\begin{align}
& p\wt{u}{}^{(0,0)}-\wt{u}{}^{(0,1)}=pu^{(0,0)}+u^{(1,0)}-\wt{u}{}^{(0,0)} u^{(0,0)},\label{B2-ZZN-a}\\
& q\wh{u}{}^{(0,0)}-\wh{u}{}^{(0,1)}=qu^{(0,0)}+u^{(1,0)}-\wh{u}{}^{(0,0)} u^{(0,0)},\label{B2-ZZN-b}\\
&-\frac{P-Q}{p-q+\wh{u}{}^{(0,0)}-\wt{u}{}^{(0,0)}}=\frac{G_3(-p,-q)}{q-p}
+(p+q+b_0)(u^{(0,0)}- \wh{\wt{u}}{}^{(0,0)})\nonumber \\
& ~~~~~~~~~~~~~~~~~~~~~~~~~~~~~~~~  -u^{(0,0)}\wh{\wt{u}}{}^{(0,0)}
+\wh{\wt{u}}{}^{(1,0)}+u^{(0,1)}, \label{B2-ZZN-c}
\end{align}
\ese
where
\begin{subequations}\label{G-func}
\begin{align}
& G_3(a,b)=g_3(a)-g_3(b),\label{G3}\\
& g_3(a)=a^3+\alpha_2  a^2, \label{ga}
\end{align}
\end{subequations}
and $b_0=-\alpha_2$.
This agrees with the DLA result (Eqs.(32a,33) in
Ref.\cite{ZZN-SAM-2012}) provided that we parameterise $P,Q$ in terms
of $p,q$ as follows:
\begin{equation}\label{eq:B2-par}
P=g_3(-p),\quad Q=g_3(-q),
\end{equation}
with  $b_0=-\alpha_2$.
Note that one can always replace $-p$ and $-q$ by $a$ and $b$,
respectively, and then get the parametrisation used in \cite{HZ-SIGMA-2011}
for getting soliton solutions.

The above parametrization of $P,Q$ provides also a connection to the
lattice potential KdV equation \cite{NQC-PLA-1983,NC-AAM-1995}: Using
(\ref{eq:B2-par}, \ref{ga}) and then taking the singular limit
$b_0\to\infty$ we get from \eqref{B2-H-c}
\begin{equation}
(x-\th x)(\t x-\h x)=-p^2+q^2.
\label{lpkdv}
\end{equation}
This one-component equation appears as H1 equation in the ABS
list \cite{ABS-CMP-2003}.

Here is a second alternative form of \eqref{B2-H} in the special case
$b_0=0$:
\bse\label{B2-W}
\begin{align}
&\omega(\wh{u}{}^{(0,1)}-\wt{u}{}^{(0,1)})=p\t u{}^{(0,0)}\!-q \h u{}^{(0,0)}
\!-u^{(0,0)}(p-q +\! \h u{}^{(0,0)}-\t u{}^{(0,0)}),~ (\omega=e^{\frac{2\pi i}{3}}),\label{B2-W-a}\\
&\h u{}^{(1,0)}-\t u{}^{(1,0)}=q \t u{}^{(0,0)}-p\h u{}^{(0,0)}+\th u{}^{(0,0)}
(p-q+\h u{}^{(0,0)}-\t u{}^{(0,0)}),\label{B2-W-b}\\
&\th u{}^{(1,0)}\!-\omega u{}^{(0,1)}=pq-(p+q+u^{(0,0)})(p+q-\th u{}^{(0,0)})
+\frac{p^3-q^3}{p-q+\h u{}^{(0,0)}-\t u^{(0,0)}}. \label{B2-W-c}
\end{align}
\ese
This is Eq.(5.3.12) in Ref.\cite{W-PhD-2001} as well as Eqs.(2.15a-c)
with $N=3$ in Ref.\cite{NPCQ-IP-1992}.  In the above
system, \eqref{B2-W-a} and \eqref{B2-W-b} can be obtained through
$\eqref{B2-ZZN-a}- \eqref{B2-ZZN-b}$ and
$\eqref{B2-ZZN-a}\h{}- \eqref{B2-ZZN-b}\t{}$, and \eqref{B2-W-c}
is \eqref{B2-ZZN-c} with $b_0=0$, in addition, $u^{(0,1)}\to -\omega
u^{(0,1)}$.

Note that the alternative forms \eqref{B2-ZZN} with \eqref{G-func}
and \eqref{B2-W} have the ``background solution'' $u^{(k,l)}=0$,
corresponding to \eqref{0ss-B2} for \eqref{B2-H}. Thus these
alternative forms will be useful once we start to construct soliton
solutions in Section \ref{sec-4} and continuum limits in Section
\ref{sec-3.3}.

\subsubsection{A2}\label{sec-3.1.2}

For the A2 equation \eqref{A2-H} (without the removable parameter
 $b_0$ (cf. \eqref{tr-A2-b0})) several alternative forms have been
 presented in the literature.

The form
\bse\label{A2-ZZN}
\begin{align}
& \wt{s}_a = (p+u^{(0,0)})\wt{v}_a-(p-a) v_a, \label{A2-ZZNa}\\
& \wh{s}_a = (q+u^{(0,0)})\wh{v}_a-(q-a) v_a, \label{A2-ZZNb}\\ &
(p+q-\wh{\wt{u}}{}^{(0,0)}+\frac{s_a}{v_a}-\alpha_2)(p-q+\wh{u}{}^{(0,0)}
-\wt{u}{}^{(0,0)})=p_a\frac{\wt{v}_a}{v_a}-q_a\frac{\wh{v}_a}{v_a},
 \label{A2-ZZNc}
\end{align}
\ese
was derived from direct linearisation approach (see Eq.(30) in
Ref.\cite{ZZN-SAM-2012}), here $p_a$ and $q_a$ are defined as
\begin{equation}\label{paqa}
 p_a=\frac{-G_3(-p,-a)}{p-a},~~ q_a=\frac{-G_3(-q,-a)}{p-a}.
\end{equation}

The transformation between Eq.\eqref{A2-H} and Eq.\eqref{A2-ZZN} is
given by
\be\label{tr-A2-H-ZZN}
v_a = \frac{x}{x_{a}}, ~~ u^{(0,0)}=z-z_0,~~s_a=\frac{1}{x_a}(y-v_ay_a),
\ee
where
\bse
\label{eq:xyz-0-A2}\bea
x_a &=& (p-a)^{-n}(q-a)^{-m}c_1, \\
z_0 &=& (c_3-p)n+(c_3-q)m+c_2, \\
y_a &=& x_a(z_0-c_3),
\eea\ese
and where $c_1$, $c_2$ are constants,  $c_3=\alpha_2/3$ (if $b_0=0$)  and
\begin{equation}\label{A2-PQ-DLA}
P=-G_3(-p,-a),~~ Q=-G_3(-q,-a).
\end{equation}
Note that the above $P,Q$ can be equivalently reparameterised as
\begin{equation}\label{A2-PQ}
P=G_3(p,a),~ Q=G_3(q,a)
\end{equation}
if we take  $c_3=-\alpha_2/3$ in \eqref{eq:xyz-0-A2}.
Either convention can be adopted.

Another system related to A2 was given in Eqs.(4.22a, 4.21b) of
Ref.\cite{N-OUP-1999}, i.e.
\bse
\label{A2-N-99}
\begin{align}
& \t s\,\t v =(p+u)\t v -p v,\\
& \h s\,\h v =(p+u)\h v -p v,\\
& (p+q+s-\th u)(p-q+\h u-\t u)=\frac{p^2 \t v-q^2 \h v}{v},
\end{align}
\ese
which can be derived from  \eqref{A2-ZZN} by taking
\be
s_a=s v,~u_0=u,~v_a=v,~a=0,~\alpha_2=0.
\ee

Then there is the system
\bse
\label{A2-N-92}
\begin{align}
& p-q+\h u-\t u =\frac{(p-a) \h v-(q-a)\t v}{\th v},\label{A2-N-92-a}\\
& p-q+\h s-\t s =(p-a) \frac{ v}{\t v}-(q-a)\frac{ v}{\h v},\label{A2-N-92-b}\\
& (p+q+s-\th u)(p-q+\h u-\t u)=\frac{p_a\t v - q_a \h v}{v}, \label{A2-N-92-c}
\end{align}
\ese
which is given by Eqs.(A.4a,b,c) in \cite{NPCQ-IP-1992}, and named as
 the ``Toda-MBSQ equation". In fact, \eqref{A2-N-92-a}
 and \eqref{A2-N-92-b} can be derived from \eqref{A2-ZZNa}
 and \eqref{A2-ZZNb}, by using
\be
s_a=s v,~ v_a=v,~ u^{(0,0)}=u,~ \alpha_2=0,
\ee
and then eliminating  $s$ and $u$, respectively.

Finally, eliminating $s_a$ from \eqref{A2-ZZNa} and \eqref{A2-ZZNb} yields
\begin{equation}
 p-q+\h u_0-\t u_0 =\frac{(p-a) \h v_a-(q-a)\t v_a}{\th v_a},
 \label{A2-Wb}
\end{equation}
which, together with \eqref{A2-ZZNa} and \eqref{A2-ZZNc} with $\alpha_2=0$,
gives the system $\{$(5.3.7a), (5.3.14), (5.3.15)$\}$ in Ref.\cite{W-PhD-2001}.

\subsubsection{C3}\label{sec-3.1.3}
The C3 equation is related to the following equation derived by
DLA \cite{ZZN-SAM-2012}
\begin{subequations}\label{C3-H-DLA-1}
\begin{align}
&(p-a)\,S_{a,b}-(p-b)\,\t S_{a,b} =\t v_{a}\,w_b,\label{C3-H-DLA-1-a}\\
&(q-a)\,S_{a,b}-(q-b)\,\h S_{a,b} =\h v_{a}\,w_b, \label{C3-H-5-DLA-b}\\
& v_{a}\,\th{w}_{b}
 =w_b\,\frac{\frac{p_a}{p-b}\,\h w_b\,\t v_a
           - \frac{q_a}{q-b}\,\t w_b\,\h v_a}{(p-b)\,\t w_b - (q-b)\,\h w_b}
 + \frac{G_3(-a,-b)}{(p-b)\,(q-b)}\,S_{a,b},\label{C3-H-DLA-1-c}
\end{align}
\end{subequations}
$S_{a,b}=s_{a,b}-1/(a-b)$ and the connection between the two equations is \cite{ZZN-SAM-2012}
\begin{subequations}\label{C3-tr}
\begin{align}
 & S_{a,b}= \left(\frac{p-a}{p-b}\right)^n\,
           \left(\frac{q-a}{q-b}\right)^m x,\label{C3-tr-a}\\
 & v_a=-(p-a)^n\,(q-a)^m y, \label{C3-tr-b}\\
 & w_b=(p-b)^{-n}\,(q-b)^{-m} z, \label{C3-tr-c}
\end{align}
\end{subequations}
where
\begin{align}\label{C3-PQb0}
P = -G_3(-p,-a),~~Q = -G_3(-q,-b),~~b_0 = G_3(-a,-b),
\end{align}
and $G_3(a,b)$, $p_a$ and $q_a$ are defined as before.

Note that since $\alpha_2$ is arbitrary we can replace $\alpha_2$ by
$-\alpha_2$ and thus in C3 \eqref{C3-H-1} $P, Q$ and $b_0$ can be
reparameterised as
\begin{align}\label{PQb0-C3}
P = G_3(p,a),~~Q =G_3(q,b),~~b_0 =- G_3(a,b),
\end{align}
where $G_3(a,b)$ is defined as \eqref{G-func}.

\section{Two- and one-component forms\label{S:1comp}}
So far we have discussed the BSQ equations in their three-component
forms, e.g., in terms of $x,y,z$, or $u^{(0,0)},u^{(1,0)},u^{(0,1)}$
etc.  As explained in Section \ref{S:evo}, Figure \ref{F1}, for the
discrete BSQ equations it is then necessary to give the initial values
for two components on an staircase-like configuration before the next
step in the evolution can be computed. An alternative formulation of
second order time evolution is to use only one component and give
initial values on two consecutive staircases as in Figure \ref{F2}. In
that case the equation usually involves the points within a $3\times
3$ stencil.  (Note that in the two-component case it is not necessary
that the equations are defined on a square, only that the next step in
the evolution can be calculated once values on the staircase are
given.)

\begin{figure}
\centering
{%\small
\begin{tikzpicture}[yscale=1.6,xscale=1.6]
\draw[step=1cm,gray,very thin] (-0.6,1.5) grid (4.6,6.5);
\draw[very thick] (-0.6,5) -- (0,5) -- (0,4) -- (1,4) -- (1,3) -- (2,3)
 -- (2,2) -- (3,2) -- (3,1.5);
\draw[very thin] (0.8,5.2) -- (3.2,5.2) -- (3.2,2.8) -- (0.8,2.8)
 -- (0.8,5.2);
\draw[very thick] (0,6.5) -- (0,6) -- (1,6) -- (1,5) -- (2,5)
 -- (2,4) -- (3,4) -- (3,3) -- (4,3) -- (4,2) -- (4.5,2);
\filldraw[black] (1,6) circle (0.05cm);
\filldraw[black] (0,6) circle (0.05cm);
\filldraw[black] (4,3) circle (0.05cm);
\filldraw[black] (4,2) circle (0.05cm);
\filldraw[black] (1,3) circle (0.05cm) node[below left] {$x_{0,0}$} ;
\filldraw[black] (1,4) circle (0.05cm) node[below left] {$x_{0,1}$} ;
\filldraw[black] (1,5) circle (0.05cm) node[below left] {$x_{0,2}$} ;
\filldraw[black] (2,3) circle (0.05cm) node[below left] {$x_{1,0}$} ;
\filldraw[black] (2,4) circle (0.05cm) node[below left] {$x_{1,1}$} ;
\filldraw[black] (2,5) circle (0.05cm) node[below left] {$x_{1,2}$} ;
\filldraw[black] (3,3) circle (0.05cm) node[below left] {$x_{2,0}$} ;
\filldraw[black] (3,4) circle (0.05cm) node[below left] {$x_{2,1}$} ;
\filldraw[red] (3,5) circle (0.05cm) node[below left] {$x_{2,2}$} ;
\filldraw[black] (0,4) circle (0.05cm);
\filldraw[black] (0,5) circle (0.05cm);
\filldraw[black] (2,2) circle (0.05cm);
\filldraw[black] (3,2) circle (0.05cm);
\end{tikzpicture}
}
\caption{For one-component second order time evolution initial values
 can be given on two consecutive staircases. The evolution equation
 will then be given on the indicated 9-point stencil.
\label{F2}}
\end{figure}

In this section we will derive two- and one-component forms of these
equations. The process of variable elimination can also be interpreted
as a B\"acklund transformation (BT). Assuming we have the following
situation:
\vspace{0.3cm}

\begin{tikzpicture}[yscale=1.5,xscale=1.5]
\node[left] at (0,0) (eq1) {$G[x]=0$};
\node[above] at (1.2,0) {eliminate $y$};
\draw [very thick, <-] (0.3,0) -- (2.1,0);
\node[right] at (2.3,0) (eq2)
{$\left\{\begin{array}{c}A[x,y]=0\\B[x,y]=0\end{array}\right.$};
\draw [very thick, ->] (4.1,0) -- (5.8,0);
\node[above] at (5,0) {eliminate $x$ \phantom{$y$}};
\node[right] at (6.1,0) (eq3) {$H[y]=0$};
\end{tikzpicture}

\vspace{0.3cm}
\noindent Then we say that the pair $\{A[x,y]=0,B[x,y]=0\}$ provides a BT
between $G[x]=0$ and $H[y]=0$.
Another way of looking at the above is to consider the pair
$\{A[x,y]=0,B[x,y]=0\}$ as two equations for one variable $x$, which
can be solved, provided that the other variable $y$ satisfies some
``integrability condition''.  In the context of PDE's this is a
familiar situation. For example, from the pair $\partial_x
\psi=A(x,y),\,\partial_y \psi=B(x,y)$ we can solve $\psi$ only if
$\partial_x B=\partial_y A$. In more general cases this problem of
``formal integrability'' or ``involutivity'' of a set of PDE's can
become quite complicated,\footnote{ The idea is to compute
  differential consequences of the initial equations and try to find
  an ``involutive completion'' after which the new differential
  consequences are just prolongations and produce no genuine new
  equations. One of the difficult problems is to decide when one can
  stop this process (this is apparently based on Spencer cohomology).}
and the problem has been analysed at length in the mathematics literature (the
reader is referred to a recent monograph \cite{seiler}).

Now we have a partial {\em difference} version of the same thing: we
can integrate one of the variables, say $x$ if the other variable, say
$y$, satisfies some \PDE (condition for integrability), which we
should find.  To do that we may need several shifted consequences of
the original equations, which provide new equations but also new
variables. The hope is that we eventually have a sufficient number of
equations to solve for all shifts of $x$ and still have one more
equation that gives a condition on the other variable.\footnote{ After
  this it remains to prove that all the remaining equations are
  satisfied due to this one condition (although this is never done in
  practice).}

 Since we will need several shifts in both directions we will use
the notation where we give as subscripts just the shifts with respect
to the basic position $(n,m)$, for example $\wt y=y_{1,0}$.

\subsection{Generalities about the elimination process}
Before studying the specific equations we can make some general
observations.  Our starting point is the set of three equations for
three variables, A2, B2, or C3. Since we always have an
equation with un-shifted $z$ it is easy to solve for that variable and
use it in the remaining ones. Similarly for $y$, we just need to apply
shifts, although due to reversal symmetry we do not have to consider
this case. Thus we can easily construct a two component pair of
equations in  $y$ and $x$ or in $z$ and $x$.

The situation with $x$ is different, because the first two equations
contain $x_{0,0}$, $x_{1,0}$, $x_{0,1}$. We can use these to eliminate
all shifted $x$ and the resulting equation would then be a polynomial
in $x_{0,0}$. In the optimal case the $x_{0,0}$-dependent part would
factor out, but this does not always happen. Furthermore it turns out
that sometimes it is beneficial to absorb some $x$-dependence into $y$
by writing the equation in terms of $w:=y/x$.

As discussed in Section \ref{S:rev-sym}, all of our equations are
reversal symmetric, which in particular exchanges $z$ and $y$. We will
therefore only need to construct two-component forms in terms of $(y,\, z)$
and $(y,\,x)$ or $(z,\, x)$.

\subsection{A2}
We take the A2 equation in the alternate form \eqref{A2-Hw} with
$w:=y/x$: \bse\label{A2-H-orig}
\begin{align}
z_{0,0}-w_{1,0}=&\frac{x_{0,0}}{x_{1,0}},\label{A2-H-a-orig}\\
z_{0,0}-w_{0,1}=&\frac{x_{0,0}}{x_{0,1}},\label{A2-H-b-orig}\\
(z_{1,0}-z_{0,1})(z_{1,1} - w_{0,0})=&P\,\frac{x_{1,0}}{x_{0,0}}
-Q\,\frac{x_{0,1}}{x_{0,0}},
\label{A2-H-c-orig}
\end{align}
\ese
because it is reversal symmetric, and because in this form $x$ appears
homogeneously and is therefore easier to eliminate.

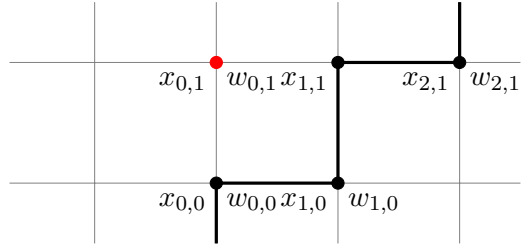
\begin{figure}
\centering
{\small
\begin{tikzpicture}[yscale=1.6,xscale=1.6]
\draw[step=1cm,gray,very thin] (-0.7,2.5) grid (3.5,4.5);
\draw[very thick] (3,4.5) -- (3,4) -- (2,4) -- (2,3) -- (1,3)
 -- (1,2.5);
\filldraw[red] (1,4) circle (0.05cm);
\draw[black] (1,4) node[below right] {$w_{0,1}$} ;
\draw[black] (1,4) node[below left] {$x_{0,1}$} ;
\filldraw[black] (1,3) circle (0.05cm) node[below left] {$x_{0,0}$} ;
\draw[black] (1,3) node[below right] {$w_{0,0}$} ;
\filldraw[black] (2,3) circle (0.05cm) node[below left] {$x_{1,0}$} ;
\draw[black] (2,3) node[below right] {$w_{1,0}$} ;
\filldraw[black] (2,4) circle (0.05cm) node[below left] {$x_{1,1}$} ;
\filldraw[black] (3,4) circle (0.05cm) node[below left] {$x_{2,1}$} ;
\draw[black] (3,4) node[below right] {$w_{2,1}$} ;
\end{tikzpicture}
}
\caption{Equations \eqref{eq:A2-xy1} and \eqref{eq:A2-xy3} allow
  evolution in the NW direction, even if initial values outside the
  square are needed.\label{F3}}
\end{figure}

\paragraph{A2 in terms of $x$ and $w:=y/x$:}
Solving for $z_{0,0}$ from \eqref{A2-H-a-orig} we get from
\eqref{A2-H-b-orig} and \eqref{A2-H-c-orig} \bse\label{eq:A2-xy}
\begin{align}\label{eq:A2-xy1}
w_{1,0}-w_{0,1}
=&\frac{x_{0,0}}{x_{0,1}}-\frac{x_{0,0}}{x_{1,0}},\\
w_{2,1}-w_{0,0}
=&\frac{x_{1,1}}{x_{0,0}}\,\frac{P\,x_{1,0} - Q\,x_{0,1}}{x_{1,0} -
 x_{0,1}}
-\frac{x_{1,1}}{x_{2,1}}.\label{eq:A2-xy2}
\end{align}
This is suitable for next eliminating $w$, but if we want to eliminate
$x$ then it is best to first eliminate $x_{0,0}$
from \eqref{eq:A2-xy2} using \eqref{eq:A2-xy1} which gives the
alternative form
\begin{equation}\label{eq:A2-xy3}
\left(w_{2,1}-w_{0,0}+\frac{x_{1,1}}{x_{2,1}}\right)(w_{1,0}-w_{0,1})
=P\,\frac{x_{1,1}}{x_{0,1}}
-Q\,\frac{x_{1,1}}{x_{1,0}}.
\end{equation}
\ese The corresponding equation for $z,x$ can be obtained by reversal
symmetry, together with \eqref{eq:rev-var-A2w}.  Note that equations
\eqref{eq:A2-xy2} and \eqref{eq:A2-xy3} are not defined on the basic
square but contain an extra point at $(2,1)$. Nevertheless they allow
evolution from a staircase initial conditions, as shown in Figure \ref{F3}.

\paragraph{A2 in terms of $z$ and $w:=y/x$:}
In order to eliminate $x$-dependence from \eqref{A2-H-a-orig} and
\eqref{A2-H-b-orig} we need to take shifts and ratios, leading to
\cite{NongThesis,FX-13}
\bse\label{eq:A2-yz}
\begin{align}
\frac{w_{1,0}-z_{0,0}}{w_{0,1}-z_{0,0}}=&
\frac{w_{1,1}-z_{0,1}}{w_{1,1}-z_{1,0}}, \label{eq:A2-yz-a}
\\
(w_{0,0}-z_{1,1})(z_{1,0}-z_{0,1})=&
\frac{P}{w_{1,0}-z_{0,0}}-\frac{Q}{w_{0,1}-z_{0,0}}.
\label{eq:A2-yz-b}
\end{align}
The second equation has the alternative form obtained by eliminating
the $z_{0,0}$-dependency
\begin{equation}
(w_{0,0}-z_{1,1})(w_{1,0}-w_{0,1})=\frac{P}{w_{1,1}-z_{0,1}}
-\frac{Q}{w_{1,1}-z_{1,0}}.\label{eq:A2-yz-c}
\end{equation}
\ese
Equations \eqref{eq:A2-yz-b} and \eqref{eq:A2-yz-c} are  connected by
  \[(P,Q,n,m,w,z) \to (-P,-Q,-n,-m,z,w).\]

The equations \eqref{eq:A2-yz} are defined on the basic quadrilateral
and are 3-dimensionally consistent; the triply shifted quantities are
given in the Appendix.

\paragraph{A2 in terms of $x$ only from \eqref{eq:A2-xy1} and \eqref{eq:A2-xy2}:}
From equations \eqref{eq:A2-xy1},\eqref{eq:A2-xy2} it is easy to
derive an equation for $x$ alone. These equations are of the following
type
\bse\label{eq:chase0}
\begin{equation}\label{eq:chase0-e}
w_{1,0}-w_{0,1}=A_{0,0},\quad w_{2,1}-w_{0,0}=B_{0,0},
\end{equation}
and by taking suitable shifts one can eliminate $w$ and derive the
``integrability condition''
\begin{equation}\label{eq:chase0-r}
B_{1,0}-B_{0,1}=A_{2,1}-A_{0,0}.
\end{equation}
\ese When this is calculated we get an equation on a $3\times 3$
stencil (cf. Figure \ref{F2}) (see Eq.(A.5) in \cite{NPCQ-IP-1992},
(5.2) in \cite{N-AAIS-1997}, Eq.(4.9) in \cite{N-OUP-1999} and
Eq.(5.7.6) in \cite{W-PhD-2001}).
\begin{equation}
 \left(\frac{P\, x_{1,1}-Q\, x_{0,2}}{x_{0,2}-x_{1,1}}\right)
  \frac{x_{1,2}}{x_{0,1}}
-\left(\frac{P\, x_{2,0}-Q\, x_{1,1}}{x_{1,1}-x_{2,0}}\right)
\frac{x_{2,1}}{x_{1,0}}
=\frac{x_{0,0}}{x_{1,0}}-\frac{x_{0,0}}{x_{0,1}}-\frac{x_{1,2}}{x_{2,2}}+
\frac{x_{2,1}}{x_{2,2}}.\label{eq:A2-x-only}
\end{equation}
This equation is reversal symmetric with \eqref{eq:rev-var-A2w} and
only changes sign.

\paragraph{A2 in terms of $w$ only from \eqref{eq:A2-yz-a}
and \eqref{eq:A2-yz-c}:} In order to derive other one component
equations one needs a slightly more complicated sequence of
elimination steps. In order to eliminate $z$ from \eqref{eq:A2-yz-a}
and \eqref{eq:A2-yz-c} we observe that the $z_{n,m}$ that appear in
these equations are located on the lattice as given in
Figure \ref{F:chase1}.1, where (a) corresponds to \eqref{eq:A2-yz-c}
and (b) to \eqref{eq:A2-yz-a}.
\begin{figure}\begin{center}
\begin{tikzpicture}[scale=0.8]
\draw[step=1cm,gray,thin] (-0.5,0.5) grid (6.5,4.5);
\fill[black] (2,2) circle (0.1cm);
\fill[black] (2,3) circle (0.1cm);
\fill[black] (1,3) circle (0.1cm);
\fill[black] (4,2) circle (0.1cm);
\fill[black] (4,3) circle (0.1cm);
\fill[black] (5,2) circle (0.1cm);
\draw (1.5,2.5) node {(a)};
\draw (4.5,2.5) node {(b)};
\draw (3,-0.5) node {Figure \ref{F:chase1}.1};
\end{tikzpicture}\hspace{2cm}\begin{tikzpicture}[scale=0.8]
\draw[step=1cm,gray,thin] (-0.5,-0.5) grid (4.5,4.5);
\draw (1,2) circle (0.25cm);
\draw (2,2) circle (0.25cm);
\draw (1,2) node {1};
\draw (2,2) node {2};
\draw (1,3) node {3};
\draw (2,3) node {4};
\draw (2,1) node {6};
\draw (3,2) node {5};
\draw (3,1) node {7};
\draw[->] (1,2) -- (1,4.6);
\draw[->] (1,2) -- (4.6,2);
\draw (2,-1.5) node {Figure \ref{F:chase1}.2};
\end{tikzpicture}
\end{center}
\caption{If equations (a) and (b) have variables located in the
lattice as in Figure \ref{F:chase1}.1 then in the elimination
process one needs to consider 7 points as given in
Figure \ref{F:chase1}.2. \label{F:chase1} }
\end{figure}

The elimination process then goes on as follows: Assume that the
$z$-values at points 1 and 2 of Figure \ref{F:chase1}.2 are arbitrary
(say, $z_{0,0}$ and $z_{1,0}$), then using (b) we can compute the
value at point 3; we denote this process by
$(1;2)\overset{(b)}{\longrightarrow} 3$. There are two routes to
compute the $z$-value at point 5:
\begin{align*}
&(1;2)\overset{(b)}{\longrightarrow} 3,\quad
(2;3)\overset{(a)}{\longrightarrow} 4,\quad
(4;2)\overset{(b)}{\longrightarrow} 5,\\
&(1;2)\overset{(a)}{\longrightarrow} 6,\quad
(2;6)\overset{(b)}{\longrightarrow} 7,\quad
(2;7)\overset{(a)}{\longrightarrow} 5.
\end{align*}
The resulting two values for $z$ at 5 must be the same, for the
arbitrary values of $z$ at points 1 and 2. Equating the two computed
values yields a rational expression in the arbitrary initial values
$z_{0,0},\,z_{1,0}$. In this case the numerator does not contain the
initial values and gives the equation. The necessary polynomial
algebra is straightforward but tedious and is best done using a
computer algebra system (such as REDUCE\cite{Hearn} or Mathematica).
The result is (see (1.3) in
\cite{NPCQ-IP-1992}, (5.3) in \cite{N-AAIS-1997}, (4.18) in
\cite{N-OUP-1999} and Eq. (5.7.3) in \cite{W-PhD-2001}.)
\begin{equation}\label{eq:A2-z-only}
\frac{P-Q}{w_{2,0} - w_{1,1}}-\frac{P-Q}{w_{1,1} - w_{0,2}}
-(w_{2,2}-w_{0,1})(w_{2,1} - w_{1,2})
-(w_{0,0}-w_{2,1})(w_{1,0} - w_{0,1})=0.
\end{equation}
The equation for $z$ alone can be obtained by reversal symmetry and
its form is identical to \eqref{eq:A2-z-only} except for sign changes (cf.\eqref{eq:rev-var-A2w}).

\paragraph{A2 in terms of $w$ only from \eqref{eq:A2-xy1}
and  \eqref{eq:A2-xy3}:}
We could also eliminate $x$ from \eqref{eq:A2-xy} in order to obtain
an equation in $w$. In the present A2 case it is not necessary
because we did already derive the $w$ equation using another
sequence of eliminations. However, in some later cases we need this
different kind of elimination process, so we will do it here as an
exercise with guaranteed success.

First note that the $x$ variables appear in lattice positions as
illustrated Figure \ref{F:chase2}, with Figure \ref{F:chase2}.1(a)
corresponding to \eqref{eq:A2-xy1} and  Figure \ref{F:chase2}.1(b)
to \eqref{eq:A2-xy3}.
\begin{figure}\centering
\begin{tikzpicture}[scale=0.7,xscale=-1,yscale=-1]
\draw[step=1cm,gray,thin] (-0.5,0.5) grid (7.5,4.5);
\fill[black] (2,2) circle (0.1cm);
\fill[black] (2,3) circle (0.1cm);
\fill[black] (1,3) circle (0.1cm);
\fill[black] (4,2) circle (0.1cm);
\fill[black] (5,3) circle (0.1cm);
\fill[black] (5,2) circle (0.1cm);
\fill[black] (6,2) circle (0.1cm);
\draw (1.5,2.5) node {(a)};
\draw (4.5,2.5) node {(b)};
\draw (3,5.5) node {Figure \ref{F:chase2}.1};
\end{tikzpicture}\hspace{1cm}
\begin{tikzpicture}[scale=0.7,xscale=-1,yscale=-1]
\draw[step=1cm,gray,thin] (-1.5,-0.5) grid (4.5,4.5);
\draw (1,2) node {1};
\draw (1,2) circle (0.25cm);
\draw (2,2) circle (0.25cm);
\draw (3,2) circle (0.25cm);
\draw (0,2) node {6};
\draw (1,1) node {7};
\draw (2,2) node {2};
\draw (1,3) node {5};
\draw (2,3) node {4};
\draw (2,1) node {8};
\draw (3,2) node {3};
\draw (3,1) node {9};
\draw (3,5.5) node {Figure \ref{F:chase2}.2};
\end{tikzpicture}
\caption{If equations (a) and (b) have variables located in the
lattice as in Figure \ref{F:chase2}.1 then in the elimination process
one need to consider 9 points as given in
Figure \ref{F:chase2}.2. \label{F:chase2}}
\end{figure}
The elimination process is now as follows: We assume that $x$ values
at circled points 1,2,3 are given and then compute the other $x$
values as follows
\begin{align*}
(1;2;3)\overset{(b)}{\longrightarrow} 4,\quad
(2;4)\overset{(a)}{\longrightarrow} 5,\quad
(1;2;5)\overset{(b)}{\longrightarrow} 6,\quad
(6;1)\overset{(a)}{\longrightarrow} 7,&
\\
(1,2)\overset{(a)}{\longrightarrow} 8,\quad
(2;3)\overset{(a)}{\longrightarrow} 9,\quad
(2;8;9)\overset{(b)}{\longrightarrow} 7.&
\end{align*}
The two values of $x$ at point 7 must be the same, which gives us an
equation which should hold for arbitrary values of $x$ at 1,2,3. When
the computations are done the result is \eqref{eq:A2-z-only}, as
expected.

\subsection{B2}
The B2 equation is given by
\bse\label{B2-H-copy}
\begin{align}
& y_{1,0}+z_{0,0} = x_{0,0}x_{1,0}\ ,\label{B2-H-a-orig}\\
& y_{0,1}+z_{0,0} = x_{0,0}x_{0,1}\ ,\label{B2-H-b-orig}\\
& y_{0,0}+z_{1,1} = x_{0,0}x_{1,1}
+b_0(x_{1,1}-x_{0,0})+\frac{P-Q}{x_{1,0}-x_{0,1}}. \label{B2-H-c-orig}
\end{align}
\ese
Note that equations (\ref{B2-H-a-orig},\ref{B2-H-b-orig}) are
similar to  (\ref{A2-H-a-orig},\ref{A2-H-b-orig}) and therefore the
derivation of two-component forms is similar.

\paragraph{B2 in terms of $y$ and $x$:}
Solving $z_{0,0}$ from \eqref{B2-H-a-orig} and using in the other
equations yields
\bse\label{eq:B2-xy}
\begin{align}\label{eq:B2-xy1}
&y_{1,0}-y_{0,1}=x_{0,0}(x_{1,0}-x_{0,1}),\\
& y_{0,0}-y_{2,1} = (x_{0,0}-x_{2,1})x_{1,1}
+b_0(x_{1,1}-x_{0,0})+\frac{P-Q}{x_{1,0}-x_{0,1}}. \label{eq:B2-xy2}
\end{align}
\ese

Another two-component form is obtained if we replace $y$ by $w:=y+z$
and after that eliminate $z$. The result is
\bse\label{RSB2}\begin{eqnarray}
w_{0,1}-w_{1,0}+(x_{0,0} + x_{1,1}) (x_{0,1} - x_{1,0})&=&0,\\
\frac{P-Q - x_{1,0}\, w_{1,0} + x_{0,1}\, w_{0,1}}{x_{1,0}-x_{0,1}}
-w_{0,0}-w_{1,1}+b_0 (x_{1,1}-x_{0,0})
\nn\\+(x_{0,0}+x_{1,1}) (x_{1,0}+x_{0,1})+x_{0,0}\, x_{1,1}&=&0.
\end{eqnarray}\ese
A notable difference with \eqref{eq:B2-xy} is that this is defined on
an elementary quadrilateral; it has the CAC property. This form was
presented in \cite{RS2017} but its relation to B2 was left open.

\paragraph{B2 in terms of $x$ only from \eqref{eq:B2-xy1}
and \eqref{eq:B2-xy2}:} The structure of equations \eqref{eq:B2-xy1}
and \eqref{eq:B2-xy2} is as in \eqref{eq:chase0-e} so that we get the
one-component equation using \eqref{eq:chase0-r}:
\begin{align}
(P-Q)\left(\frac1{x_{2,0}-x_{1,1}}-\frac1{x_{1,1}-x_{0,2}}\right)
+b_0(x_{0,1}-x_{1,0}+x_{2,1} -x_{1,2})&\nonumber  \\
-(x_{2,2}-x_{0,1})(x_{2,1}-x_{1,2})
-(x_{0,0}-x_{2,1})(x_{1,0}-x_{0,1})&=0. \label{eq:B2-x-only}
\end{align}

This equation is a generalization ($b_0$-terms)
of \eqref{eq:A2-z-only} derived for A2.

Attempts to eliminate $x$ seem to lead into very complicated equations.

\subsection{C3}
As noted in Section \ref{SS:C34} we only need to consider C3 equation
\bse
\label{eq:CH}
\begin{eqnarray}
 y_{1,0} \,z_{0,0}&=&x_{1,0} - x_{0,0},\label{eq:CH-a}\\
 y_{0,1} \,z_{0,0}&=&x_{0,1} - x_{0,0},\label{eq:CH-b}\\
 z_{1,1}\,y_{0,0}&=& b_0\,x_{0,0} + b_1
 + z_{0,0} \frac{P\,y_{1,0}\,z_{0,1}-Q\,y_{0,1}\,z_{1,0}}
 {z_{1,0}-z_{0,1}},\label{eq:C3}
\end{eqnarray}
\ese for the three-component case $b_0\neq0,b_1=0$ and for the two
component case $b_0=0$, $b_1$ arbitrary. We can treat them together for some
computations.

\paragraph{C3 in terms of $x,y$:}
After solving for $z$ from \eqref{eq:CH-a} and using it in
\eqref{eq:CH-b} and \eqref{eq:C3} we get the two-component form
\bse\label{eq:Cxy}
\begin{align}
y_{1,0}(x_{0,1}-x_{0,0})=&y_{0,1}(x_{1,0}-x_{0,0}),\label{eq:Cxy-simple}\\
y_{0,0}(x_{2,1}-x_{1,1})=&y_{2,1}(b_0\,x_{0,0}+b_1)\nonumber\\
+&y_{2,1}\frac
{(x_{1,1}-x_{0,1})(x_{0,0}-x_{1,0})\,P-
(x_{1,1}-x_{1,0})(x_{0,0}-x_{0,1})\,Q}{x_{1,0}-x_{0,1}}.\label{eq:Cxyz-simple}
\end{align}
This form is well suited for eliminating $y$ next, because it is linear
in $y$, but if we would like to eliminate $x$ the following
alternative would be useful:
\begin{align}
(x_{2,1}-x_{1,1})\frac{y_{0,0}}{y_{2,1}}
  = (b_0\,x_{0,0}+b_1)-\frac{(x_{1,1}-x_{0,1})\,y_{1,0}\,P-
(x_{1,1}-x_{1,0})\,y_{0,1}\,Q}{y_{1,0}-y_{0,1}},\label{eq:Cxy-alt}
\end{align}
\ese
 because it is linear in $x$. These equations are defined on a
  configuration given in Figure \ref{F3} ($w\leftrightarrow y$). The
equations for $x,z$ are the same as above, and can be obtained by
reversal, cf.(\ref{eq:rev-var-C},\ref{eq:C-rev-par}).

\paragraph{C3 in terms of $x$ only:}
Equations (\ref{eq:Cxy-simple}, \ref{eq:Cxyz-simple}) have the form
\[
\frac{y_{1,0}}{y_{0,1}}=A_{0,0},\quad
\frac{y_{0,0}}{y_{2,1}}=B_{0,0}.
\]
The dependence is similar to \eqref{eq:chase0}, except that instead of
an additive case we now have a multiplicative case. The integrability
condition is
\[
\frac{A_{2,1}}{A_{0,0}}=\frac{B_{0,1}}{B_{1,0}},
\]
which yields
%\begin{align}
% &\frac{(x_{2,2}-x_{1,2})(x_{0,2}-x_{1,1})(x_{0,1}-x_{0,0})}
%{(x_{2,2}-x_{2,1})(x_{1,1}-x_{2,0})(x_{1,0}-x_{0,0})}=\nonumber\\
%&\frac{(x_{1,1}-x_{0,2})(b_0\,x_{0,1}+b_1)
%+(x_{1,2}-x_{0,2})(x_{0,1}-x_{1,1})\,P-
%(x_{1,2}-x_{1,1})(x_{0,1}-x_{0,2})\,Q}
%{(x_{2,0}-x_{1,1})(b_0\,x_{1,0}+b_1)
%+(x_{2,1}-x_{1,1})(x_{1,0}-x_{2,0})\,P-
%(x_{2,1}-x_{2,0})(x_{1,0}-x_{1,1})\,Q}.\label{eq:C34-x-only}
%\end{align}
\begin{equation}
\begin{array}{l}\displaystyle
 \frac{(x_{2,2}-x_{1,2})(x_{0,2}-x_{1,1})(x_{0,1}-x_{0,0})}
{(x_{2,2}-x_{2,1})(x_{1,1}-x_{2,0})(x_{1,0}-x_{0,0})}= \\
\displaystyle
\frac{(x_{1,1}\!-x_{0,2})(b_0\, x_{0,1}\!+b_1)
+(x_{1,2}\!-x_{0,2})(x_{0,1}-x_{1,1})P-
(x_{1,2}-x_{1,1})(x_{0,1}-x_{0,2})Q}
{(x_{2,0}\!-x_{1,1})(b_0\,x_{1,0}\!+b_1)
+(x_{2,1}\!-x_{1,1})(x_{1,0}-x_{2,0})P-
(x_{2,1}-x_{2,0})(x_{1,0}-x_{1,1})Q}.
\end{array}
\label{eq:C34-x-only}
\end{equation}
 Note that this is invariant under $n\leftrightarrow
  m,\,P\leftrightarrow Q $, and reversal symmetric with
  \eqref{eq:C-rev-par}.

\paragraph{C3 in terms of $y$ only:}
The dependence on $x$ in \eqref{eq:Cxy-simple} and \eqref{eq:Cxy-alt}
is linear and of the form given in Figure \ref{F:chase2}, and using
that method we can eliminate $x$. The resulting
equation in terms of $y$ is the same as \eqref{eq:A2-x-only} in terms
of $x$. Note in particular that dependence on the parameters $b_0,b_1$
drops out from the $y$-equation, while it remains in the $x$-equation.

\paragraph{C3 in terms of  $y$ and $z$:}
The method of eliminating $x$ depends sensitively on the additional
$x$-dependent term. We need to consider separately the two cases.

\begin{itemize}
\item $b_1$ arbitrary, $b_0=0$. In this case equation \eqref{eq:C3}
does not depend on $x$ at all, while form \eqref{eq:CH-a}
and \eqref{eq:CH-b} $x$ can be easily eliminated, leaving
\bse\label{eq:Csimple-both}
\begin{align}
y_{1,1}(z_{1,0}-z_{0,1})&+z_{0,0}(y_{1,0}-y_{0,1})=0, \label{eq:Csimple}\\
y_{0,0}z_{1,1}&=b_1 +z_{0,0}\frac{P y_{1,0}z_{0,1}-Q
y_{0,1}z_{1,0}}{z_{1,0}-z_{0,1}}.\label{eq:Csimple-2}
\end{align}
\ese
Note that by the gauge transformation
\begin{equation}
z=w\,p^n\,q^m, \quad y=-v\,p^{-n}\,q^{-m},
\label{tr-C3-3c2c}
\end{equation}
with $P=p^3$, $Q=q^3$ and $b_1=b'_1 pq$, we will get from
\eqref{eq:Csimple-both}
\begin{equation}\label{C3-N-2c-b1}
 \frac{p\,\t w-q\,\h w\,}{w}
=\frac{p\,\h v-q\,\t v\,}{\th v}
=\frac{p\,\t v\,\h w-q\,\h v\,\t w\,}{b'_1 +v\,\th w}.
\end{equation}
When $b'_1=0$ we get
\begin{equation}\label{C3-N-2c}
 \frac{p\,\t w-q\,\h w\,}{w}
=\frac{p\,\h v-q\,\t v\,}{\th v}
=\frac{p\,\t v\,\h w-q\,\h v\,\t w\,}{v\,\th w},
\end{equation}
which was already presented in \cite{N-OUP-1999}.

\item $b_1=0,\, b_0\neq 0$. In this case it is best to
absorb some $x$ into $y$ by defining $w:=y/x$. Then \eqref{eq:CH-a}
and \eqref{eq:CH-b} yield
\bse\label{eq:C3-2-yz}
\begin{equation}\label{eq:C3-2-yz-a}
w_{1,1}z_{0,0}(w_{1,0}z_{1,0}-w_{0,1}z_{0,1})-w_{1,1}(z_{1,0}-z_{0,1})
-z_{0,0}(w_{1,0}-w_{0,1})
=0,
\end{equation}
while the other equation becomes
\begin{equation}\label{eq:C3-2-yz-b}
b_0+
\frac{w_{1,1}}{w_{1,0}-w_{0,1}}
\left(\frac{P w_{1,0}z_{0,1}}{w_{1,1}z_{0,1}-1}
-\frac{Q w_{0,1}z_{1,0}}{w_{1,1}z_{1,0}-1}\right)
-w_{0,0}z_{1,1}=0.
\end{equation}
\ese
\end{itemize}
Note that in both cases the  equation pair is still quadrilateral.

In order to derive one-component forms one can use the method of Figure
\ref{F:chase1}, after some modifications in the equations.

\paragraph{C3 in terms of $z$ or $y$ from
  \eqref{eq:Csimple-both}:} Eliminating variable $y$ leads to
\eqref{eq:A2-x-only} but in terms of $z$ and $n,m$ reversed.

In order to eliminate $z$ one should first eliminate $z_{0,0}$ from
\eqref{eq:Csimple-2} using \eqref{eq:Csimple}, after which the method
of Figure \ref{F:chase1} works and yields \eqref{eq:A2-x-only} in
terms of $y$.

\paragraph{C3 in terms of $w$ or $z$ from \eqref{eq:C3-2-yz}:}
After eliminating $z$ one gets \eqref{eq:A2-x-only}, but now in terms
of $w$ and with $(p,q)\mapsto (p-b_0,p-b_0)$.

In order to eliminate $w$ one should first eliminate $w_{1,1}$ from
\eqref{eq:Csimple-2} using \eqref{eq:Csimple}, and then one gets
\eqref{eq:A2-x-only} in terms of $z$ but $n,m$ reversed.

\paragraph{Summary of one-component forms:}
\begin{itemize}
\item A2 has two one-component forms: \eqref{eq:A2-z-only} in $w=y/x$
  and in $z$, and \eqref{eq:A2-x-only} in $x$. They correspond to
  regular and modified BSQ, respectively.
\item C3 has two one-component forms: \eqref{eq:C34-x-only} in $x$ and
  \eqref{eq:A2-x-only} in $y$ or $w=y/x$ or $z$, corresponding to Schwarzian
  and modified BSQ, respectively.
  \end{itemize}
Thus we can say that the three-component form of A2 contains both
regular and modified BSQ and the three-component form of C3 contains
both Schwarzian and modified BSQ. B2 on the other hand contains only
regular BSQ, but in a generalized form in comparison to A2. It is an
interesting open question whether there is a three-component version
which contains all three different BSQ equations in full generality.

\section{Continuum limits}\label{sec-3.3}
We will now consider (semi-)continuous limits of the derived
one-component equations: \eqref{eq:A2-x-only} for A2,
\eqref{eq:B2-x-only} for B2, and \eqref{eq:C34-x-only} for
C3$_{b_0}$. These are defined on a $3\times 3$ stencil, see Figure
\ref{F:squeeze}.

The technical aspects of taking semi-continuous and fully continuous
limits involve several choices, including the gauge (or background
solution) and the way the lattice parameters behave under these
limits. For example, we know that $p,q$ are related to lattice
spacing, but should we let, for example, $p\to 0$ or $p\to\infty$?
Before going into specific equations we are going to discuss some
aspects that apply to all cases.

\subsection{Common features}
\subsubsection{Approaching continuum}
The semi-continuous limit means that in some direction the lattice
points approach each other to form a continuum.
There are two simple ways to squeeze the $3\times 3$ stencil onto a line,
the straight and the skewed way, see Figure \ref{F:squeeze}.1 and
\ref{F:squeeze}.2, respectively. (Squeezing in other directions are
also possible but the result would depend on still more points and
probably would not be useful for applications.)

\begin{figure}\centering
\begin{tikzpicture}[scale=0.8,xscale=1,yscale=1]
\draw[step=1cm,gray,thin] (1.5,1.5) grid (4.5,4.5);
\fill[black] (2,4) circle (0.1cm);
\fill[black] (2,3) circle (0.1cm);
\fill[black] (2,2) circle (0.1cm);
\fill[black] (3,4) circle (0.1cm);
\fill[black] (3,3) circle (0.1cm);
\fill[black] (3,2) circle (0.1cm);
\fill[black] (4,4) circle (0.1cm);
\fill[black] (4,3) circle (0.1cm);
\fill[black] (4,2) circle (0.1cm);

\draw[gray,thin] (0.5,0.5) -- (5.5,0.5);
\draw[->,line width=0.6pt] (2,4.5) -- (2,1.1);
\draw[->,line width=0.6pt] (3,4.5) -- (3,1.1);
\draw[->,line width=0.6pt] (4,4.5) -- (4,1.1);
\fill[black] (2,0.5) circle (0.1cm);
\fill[black] (3,0.5) circle (0.1cm);
\fill[black] (4,0.5) circle (0.1cm);

\draw (3,-1.5) node {Figure \ref{F:squeeze}.1};

\end{tikzpicture}\hspace{1cm}\begin{tikzpicture}[scale=0.8,xscale=1,yscale=1]
\draw[step=1cm,gray,thin] (1.5,1.5) grid (4.5,4.5);
\fill[black] (2,4) circle (0.1cm);
\fill[black] (2,3) circle (0.1cm);
\fill[black] (2,2) circle (0.1cm);
\fill[black] (3,4) circle (0.1cm);
\fill[black] (3,3) circle (0.1cm);
\fill[black] (3,2) circle (0.1cm);
\fill[black] (4,4) circle (0.1cm);
\fill[black] (4,3) circle (0.1cm);
\fill[black] (4,2) circle (0.1cm);

\draw[gray,thin] (4,-0.5) -- (6.5,2);

\draw[->,line width=0.6pt] (2,4) -- (4.75,1.25);
\draw[->,line width=0.6pt] (3,4) -- (5.25,1.75);
\draw[->,line width=0.6pt] (4,4) -- (5.75,2.25);
\draw[->,line width=0.6pt] (2,3) -- (4.25,0.75);
\draw[->,line width=0.6pt] (2,2) -- (3.75,0.25);

\fill[black] (4.25,-0.25) circle (0.1cm);
\fill[black] (4.75,0.25) circle (0.1cm);
\fill[black] (5.25,0.75) circle (0.1cm);
\fill[black] (5.75,1.25) circle (0.1cm);
\fill[black] (6.25,1.75) circle (0.1cm);

\draw (3,-1.5) node {Figure \ref{F:squeeze}.2};
\end{tikzpicture}
\caption{Two ways to project the $3\times 3$ stencil to a line: Figure
  \ref{F:squeeze}.1 gives the straight limit, Figure \ref{F:squeeze}.2
  the skew limit. A $90^o$ degree rotation is also possible.
 \label{F:squeeze}}
\end{figure}

Before taking the limit we must anchor the discrete and continuous
variables. First of all we shift the equations so that the center of
the $3\times 3$ stencil is at $(0,0)$. For the {\em straight
  limit} in the $m$ direction, we can take $m=0$ to correspond to
$\xi=0$ and then generically the discrete variable $m$ corresponds to
continuous variable $\xi$ and we can write
\begin{equation}\label{eq:str-def}
u_{n,\mu}=u_n(\xi+\epsilon \mu),
\end{equation}
where $\epsilon$ measures the distance between two lattice points in
the $m$ direction. In practice $\mu\in\{-1,0,1\}$.
For the {\em skew limit} we must take instead
\begin{equation}\label{eq:skw-def}
x_{\nu,\mu}=u_{\nu+\mu}(\tau+\tfrac12(\nu-\mu)\epsilon),
\end{equation}
where $\nu,\mu\in\{-1,0,1\}$.

Now that the dependence on $\epsilon$ is given we can expand
in $\epsilon$,
\begin{equation}\label{eq:str-exp}
u_n(\xi+\epsilon s)=u_n(\xi)+\epsilon s\, u_n'(\xi)+\frac12\epsilon^2 s^2\,
u_n''(\xi)+\cdots.
\end{equation}

\subsubsection{Parameter relations}
The above discussion was about limits in general, but in concrete
cases we must first determine the connections between the various
parameters in the equation and the lattice spacing. One method to get
information about this is to linearize the discrete equation and study
its discrete plane wave solutions or plane wave factors (PWF)
(cf.\cite{QNCL,NC-AAM-1995}) Once the PWF is known one can figure out
what should be done in order to get, as a limit, the continuous
PWF. This is described in Chapter 5 of \cite{HJN}. In general terms
the discrete-continuous relation is an application of
\begin{equation}\label{dclim}
\lim_{n\to\infty}\left(1+\frac{\alpha}{n}\right)^n=e^\alpha.
\end{equation}

Before one can linearize a nonlinear equation it is necessary to
choose a background solution (or gauge) around which to expand. For
equation \eqref{eq:A2-x-only}
we must  choose a multiplicative gauge
\begin{equation}\label{eq:A2-gauge}
x_{n,m}=(p-a)^{-n+n_0}(q-b)^{-m+m_0}c_1(1+\epsilon v_{n,m}+\cdots),
\end{equation}
while for \eqref{eq:B2-x-only} we must choose additive gauge, such as
\begin{equation}\label{eq:B2-gauge}
x_{n,m}=\sigma (n-n_0) (p-a)+\sigma (m-m_0) (q-b)+\epsilon u_{n,m}+\cdots,
\end{equation}
but for  \eqref{eq:C34-x-only} no gauge is necessary.

We have used here $(n_0,m_0)$ as origin but in the expansion these
should drop out.

For each of the considered DBSQ equations one finds
for $u_{n,m}$ or $v_{n,m}$  the PWF solution:
\begin{flalign}
  \hbox{PWF:}\hspace{4cm}& \biggl(\frac{p-\omega(k)}{p-k}\biggr)^n
  \biggl(\frac{q-\omega(k)}{q-k}\biggr)^m,&
\label{PWF}
\end{flalign}
where $\omega(k)\neq k$ is one of the roots of $g_3(\omega)-g_3(k)=0$,
where  $g_3$ is a cubic polynomial, such as \eqref{ga}.

\paragraph{Straight limit:}
In order to use \eqref{dclim} for a straight limit in the
$m$-direction we write
\[
\biggl(\frac{q-\omega(k)}{q-k}\biggr)^m=
\biggl(1+\frac{k-\omega(k)}{q-k}\biggr)^m
\]
and therefore $q$ must approach infinity as $m$, i.e., $m/q=\xi$. Then
\[
\biggl(\frac{q-\omega(k)}{q-k}\biggr)^m=
\biggl(1+\frac{\xi(k-\omega(k))}{m-\xi k}\biggr)^m\!\longrightarrow ~
\exp[\xi(k-\omega(k))],\quad\text{as }m\to\infty.
\]
Sometimes we may need higher order corrections to the above as follows:
\[
\biggl(\frac{q-\omega(k)}{q-k}\biggr)^m=
\exp\left[\xi(k-\omega(k))+\tfrac12(k^2-\omega(k)^2)\xi/q+O (1/q^2)\right].
\]
These results suggest that in a straight limit in $m$ direction we should take
\begin{equation}\label{eq:str-par}
m=\xi\,q, \quad q\to\infty.
\end{equation}

\paragraph{Skew limit:}
In the skew limit, as in Figure \ref{F:squeeze}.2, we first need to
make a $45^o$ degree rotation, and after defining
$N=n+m,\,m'=\tfrac12(n-m)$, i.e., $n=\tfrac12N+m',\,m=\tfrac12N-m'$,
we get\footnote{Note that one can also take $N=n+m,\,m'=m$, cf.\cite{HJN}.}
\begin{align*}
&\biggl(\frac{p-\omega(k)}{p-k}\biggr)^n\biggl(\frac{q-\omega(k)}{q-k}\biggr)^m\\
=~&\biggl(\frac{p-\omega(k)}{p-k}\cdot\frac{q-\omega(k)}{q-k}\biggr)^{N/2}
\biggl(\frac{p-k}{p-\omega(k)}\cdot \frac{q-\omega(k)}{q-k}\biggr)^{m'}\\
  =~&\biggl(\frac{p-\omega(k)}{p-k}\cdot\frac{q-\omega(k)}{q-k}\biggr)^{N/2}
  \left(1+\frac{(k-\omega(k))(p-q)}{(p-\omega(k))(q-k)}\right)^{m'}.
  \label{PWF-sk0}
\end{align*}
This suggests that $p-q$ should approach 0, thus we take
\begin{equation}\label{eq:skw-par}
q=p-\delta,\quad \tau=\delta m',\quad \delta\to 0
\end{equation}
and then the expansion yields
\[
\left(1+\frac{(k-\omega(k))\delta}
     {(p-\omega(k))(p-k-\delta)}\right)^{\tau/\delta}
     =\exp\left[\frac{(k-\omega(k))\tau}{(p-\omega(k))(p-k)}
       +O(\delta^2)\right].
\]

Thus the main results we have so far obtained from PWF analysis are the
limit behaviors \eqref{eq:str-par} and \eqref{eq:skw-par} for straight
and skew limits, respectively. Expanding in higher orders will be
useful in deriving double limits.

Finally we must relate the lattice parameters $p,q$
used in the limits to the parameters $P,Q$ appearing in the equation,
we use
\begin{equation}\label{eq:PQpq}
P=g_3(p)\text{ or }P=g_3(p)-g_3(a),\, \text{ where }\,
  g_3(x):=x^3+\alpha_2 x^2+\alpha_1 x+\alpha_0,\quad
\end{equation}
and similarly for $Q$. (This $g_3$ is a generalized version of
\eqref{ga} with additional $\alpha_j$.) The cubic term is a signature
of Boussinesq equations. It should also be noted that as far as the
CAC test is concerned, the parameters $P,Q$ can be arbitrary. However,
if we want the discrete equation to have a continuum limit to a
(semi-)continuous Boussinesq equation then the cubic form is
necessary, furthermore $P,Q$ may depend on some other parameters of
the equation.

\subsection{B2}\label{sec-3.3.2}
  As the first case we take equation
\eqref{eq:B2-x-only}, but centered at the origin
\begin{align*}
(P-Q)\left(\frac1{x_{1,-1}-x_{0,0}}-\frac1{x_{0,0}-x_{-1,1}}\right)
+b_0(x_{-1,0}-x_{0,-1}+x_{1,0} -x_{0,1})&  \\
-(x_{1,1}-x_{-1,0})(x_{1,0}-x_{0,1})
-(x_{-1,-1}-x_{1,0})(x_{0,-1}-x_{-1,0})&=0.
\end{align*}
\subsubsection{Straight limit}\label{b2-str}
We use the linear background $x=pn+qm+\epsilon\, u$ and take
$\epsilon=1/q\to 0$ with initially arbitrary $P,Q$ as in
\eqref{eq:PQpq}. Then using \eqref{eq:str-def} and expanding
we find from order 0 that $\alpha_3=1$, and at order 1
that $\alpha_2=-b_0$.\footnote{ This means in particular that although
  $P,Q$ and $b_0$ were independent in the CAC analysis, they must be
  related in the indicated way before we have a reasonable continuum
  limit.}  Finally (if $\alpha_1=0$, in agreement with \eqref{ga}) at
order 2 we get the semi-discrete three-point equation
\begin{align}
&u_1''+u_0''+u_{-1}''
-3u_1'(p-u_0+u_1)
+3u_{-1}'(p-u_{-1}+u_0)\nn\\
&+(p-u_0 +u_1)^3-(p- u_{-1}+u_0)^3\nn\\
&-b_0[u'_{-1}-u'_{1}+(2p-u_{-1}+u_1)(u_{-1}-2u_0+u_1)]=0.
\label{B2-CL-st1}
\end{align}
where the primes refer to derivatives with respect to $\xi$,
cf.~\eqref{eq:str-def}.  Note that when $b_0=0$, the above equation is
Eq.(5.9.13) in \cite{W-PhD-2001} (with $u \to -u$).  The $b_0$ term is
actually a discrete derivative of the semi-discrete lpKdV equation
\[
u'_{0}+u'_{1}+(2p-u_{0}+u_1)(u_{0}-u_1)=0.
\]

\subsubsection{Skew limit}
 As discussed before, we should make a $45^o$ degree rotation and then
take the continuous limit in the $m'$ direction.   We use the same
linear background as before, and the form of $P,Q$ found for the
straight limit, but instead of \eqref{eq:str-par} and
\eqref{eq:str-def} we use \eqref{eq:skw-par} and \eqref{eq:skw-def}.

Then at order $\epsilon^3$ we get the following five-point equation
\begin{align}
(3p^2-2 p b_0)\,\ddot u_0
+(\dot u_0+1)^2[&(u_2-u_{-1}-b_0+3p)(\dot u_1+1)\nn\\
&+(u_{-2}-u_{1}+b_0-3p)(\dot u_{-1}+1)]=0,\label{eq:B2skew}
\end{align}
where the dot refers to derivatives with respect to $\tau$ in
\eqref{eq:skw-def}.  For $b_0=0$ this agrees with (5.9.4) in
\cite{W-PhD-2001}, up to $p\to -p$.

\subsubsection{Double limit}
Now we take equation \eqref{B2-CL-st1} and do the straight continuum
limit on the remaining discrete variable using $p=1/\delta$. We set
$u_{\nu}(\xi)=v(\tau+\nu\delta,\xi)$ and apply Taylor expansion
as in \eqref{eq:str-exp} and the leading term in the expansion would
then be $(\partial_\xi -\partial_\tau )v(\tau ,\xi )=0$, which means
that the naive limit does not work. In order to get a better
understanding of the situation we return to the PWF \eqref{PWF}. With
$q\to\infty$ limit already taken we expand in $1/p$:
\begin{equation}\label{PWF-st2}
\biggl(\frac{p-\omega(k)}{p-k}\biggr)^n\! e^{(k-\omega(k))\xi }
=\exp\!\left[{(k\!-\omega(k))(\xi\!+\tau)+ \tfrac{1}{2} (k^2\!-\omega^2(k))
  \tfrac{\tau}{p} +O(\tfrac{1}{p^2})}\right],~ \tau=\frac{n}{p}.
\end{equation}
This suggests that we should introduce new variables
\begin{equation}
x=\xi+\tau,~~ t=\tau/p,
\end{equation}
which means
\begin{equation}\label{eq:B2-dch}
\partial_\xi=\partial_x,~~\partial_\tau=\partial_x+\tfrac{1}{p}\partial_t,
\end{equation}
and apply these in the expansion, before actually taking the limit
$p\to\infty$. In other words, the variables $\xi $ and $\tau $ are
infinitesimally close and therefore the leading term combines with
some higher order terms.  With this modification the leading term
becomes
\[
v_{xxxx}-12 v_{xx}v_x -8b_0v_{xt}+12v_{tt}=0.
\]
The $v_{xt}$ term can be converted into $v_{xx}$ by the translation
$\partial_t\mapsto
\partial_t+\tfrac13b_0 \partial_x$ (which could have been be added into
\eqref{eq:B2-dch}) which yields
\begin{equation}\label{eq:B2-dbl}
v_{xxxx}-12 v_{xx}v_x-\tfrac43b_0^2v_{xx}+12v_{tt}=0.
\end{equation}
This is the standard Boussinesq equation with a mass term.

We can also take the double continuum limit by starting from
\eqref{eq:B2skew}, where the continuous variable is $\tau$ and
$p=1/\delta$. Thus we expand $u_{\nu}(\tau)=v(\xi+\nu\delta,\tau)$,
but a change of the continuous variables is also necessary.
If we define $x,t$ by
\[
\partial_\tau=[1+\delta\tfrac13b_0]\delta^2\partial_x
+\delta^3\partial_t,\quad\partial_\xi=\partial_x,
\]
we get \eqref{eq:B2-dbl}.

Finally, we can take a double limit directly from the fully discrete
form \eqref{eq:B2-x-only}. For that purpose we take the linear gauge
as before, and the limit by
$p=p_0/\epsilon,\,q=q_0/\epsilon,\,\epsilon\to0$ with
\begin{equation}\label{eq:dbl-limi}
u_{n,m}=-v(x+n A+m B, y+(n+m)\epsilon^2/(p_0q_0)),
\end{equation}
where
\begin{equation}\label{eq:dbl-limi-B2}
A:=-\epsilon/p_0-(\epsilon^2 b_0)/(3 p_0^2),\quad
B:=-\epsilon/q_0-(\epsilon^2 b_0)/(3 q_0^2).
\end{equation}
After  expanding in $\epsilon$ the leading term is again \eqref{eq:B2-dbl}.

\subsection{A2}\label{sec-3.3.3}
Now we consider equation \eqref{eq:A2-x-only}, centered at the origin:
\begin{equation*}
 \left(\!\frac{P\, x_{0,0}-Q\, x_{-1,1}}{x_{-1,1}-x_{0,0}}\!\right)
 \! \frac{x_{0,1}}{x_{-1,0}}
-\left(\!\frac{P\, x_{1,-1}-Q\, x_{0,0}}{x_{0,0}-x_{1-1}}\!\right)
\!\frac{x_{1,0}}{x_{0,-1}}
=\frac{x_{-1,-1}}{x_{0,-1}}-\frac{x_{-1,-1}}{x_{-1,0}}-\frac{x_{0,1}}{x_{1,1}}+
\frac{x_{1,0}}{x_{1,1}}.
\end{equation*}
In this case we use the gauge transformation \eqref{eq:A2-gauge} and
the cubic parameterization \eqref{eq:PQpq}.

\subsubsection{Straight limit}
As in Section \ref{b2-str} we use \eqref{eq:str-par}, \eqref{eq:PQpq} and
\eqref{eq:str-def} but now with a multiplicative gauge
\eqref{eq:A2-gauge} and take $\alpha_3=\sigma=1,\,b=a$ after which the
leading term in the expansion will yield a second order three-point equation:
\begin{align}
\frac{v_1''}{v_1}+&\frac{v_0''}{v_0}+\frac{v_{-1}''}{v_{-1}}
-2\left(\frac{v_{1}'}{v_{1}}\right)^2
-\left(\frac{v_{0}'}{v_{0}}\right)^2
-\frac{v_1'v_{0}'}{v_1v_{0}}
+\frac{v_{-1}'v_{0}'}{v_{-1}v_{0}}\nonumber\\
+&3(p-a)\left(\frac{v_{-1}'}{v_0}-\frac{v_1'v_0}{v_1^2}\right)
+(\alpha_2+3a)\left(\frac{v_{-1}'}{v_{-1}}-\frac{v_{1}'}{v_{1}}\right)
-\left(\frac{v_{1}}{v_{0}}-\frac{v_{0}}{v_{-1}}\right)
\frac{P(p)}{a-p}\nonumber\\
-&\left(\frac{v_{-1}}{v_{0}}-\frac{v_{0}}{v_{1}}\right)
(\alpha_2+3a)(a-p)
+\left(\frac{v_{-1}^2}{v_{0}^2}-\frac{v_{0}^2}{v_{1}^2}\right)(a-p)^2=0.
\label{eq:A2-str}
\end{align}
This can also be written as
\begin{align}
[\ln(v_{-1}v_0v_1)]''+&(\ln v_{-1}-\ln v_1)'[\alpha_2+3a+(\ln
  (v_{-1}v_0v_1))'] \nn
\\ +&(p-a)\left[\frac{v_{-1}}{v_0}(\alpha_2+3a+3(\ln v_{-1})')
  -\frac{v_{0}}{v_1}(\alpha_2+3a+3(\ln
  v_{1})')\right]\nn\\ +&\frac{P(p)}{p-a}\left(\frac{v_{1}}{v_{0}}
-\frac{v_{0}}{v_{-1}}\right)
+(p-a)^2\left(\frac{v_{-1}^2}{v_{0}^2}-\frac{v_{0}^2}{v_{1}^2}\right)=0.
\label{eq:A2-stra}
\end{align}
If  $\alpha_2=a=0$ this is the same as (5.9.14) in \cite{W-PhD-2001}.

\subsubsection{Skew limit}
For the skew limit we also use multiplicative gauge \eqref{eq:A2-gauge}
and expansion \eqref{eq:skw-def} with \eqref{eq:skw-par} and cubic
parameterization. The semi-continuous limit is then
\begin{equation}\label{eq:A2-skw}
\partial_\tau\left(\frac{v_1}{v_{-1}}\,
\frac{\Pi_1\,\dot v_0+\Pi_2\,v_0}{(p-a)\dot v_0-v_0}\right)
-\frac{v_{-2}}{v_{-1}}\left((p-a)\frac{\dot v_{-1}}{v_{-1}}-1\right)
+\frac{v_{1}}{v_{2}}\left((p-a)\frac{\dot v_{1}}{v_{1}}-1\right)=0,
\end{equation}
where
\begin{eqnarray*}
\Pi_1&=&(g_3(p)-g_3(a))/(a-p),\\
\Pi_2&=&-(a+2p+\alpha_2).
\end{eqnarray*}

\subsubsection{Double limit}
Again we can derive the fully continuous limit from either the
straight semi-continuous equation \eqref{eq:A2-str} or from the skew
semi-continuous limit \eqref{eq:A2-skw} or taking a double limit
directly from \eqref{eq:A2-x-only}. For A2 there is the special
feature that in order to reach the modified BSQ equation we must also
change the dependent variables by $v=e^V$.

Starting with the straight semi-continuous limit \eqref{eq:A2-str} we
use again $p=1/\delta$ and $u_\nu=v(y+\nu\delta,x),\,v=e^V$ and
furthermore we use the change of variables\footnote{This
  transformation is needed in order to eliminate the cross term $V_{xt}$ in
  favor of the mass term $V_{xx}$. }
\[
\partial_y=(1-\tfrac13\alpha_2\delta)\partial_x+\tfrac12\delta\partial_t,
\]
and then we obtain
\begin{equation}\label{eq_A2-dbl}
V_{xxxx}-6 V_{xx} V_x^2+6V_{xx}V_t-4(\alpha_2+3a)V_{xx}V_x
+4(\alpha_1-\tfrac13\alpha_2^2) V_{xx} +3V_{tt}=0.
\end{equation}
This is the modified Boussinesq equation (cf. \cite{HirS-PTP-1977},
Equation (3.5)).

From the skew limit \eqref{eq:A2-skw} we use
$p=1/\delta$, $\alpha_3=1$ and the variable change
\[
\partial_y=\delta^2(-1+\tfrac13\alpha_2\delta)\partial_x
-\delta^3\partial_t,
\]
and get the same result \eqref{eq_A2-dbl}.

Finally the double limit directly from
\eqref{eq:A2-x-only} is taken with \eqref{eq:dbl-limi} except now
\begin{equation}\label{eq:dbl-limi-A2}
A:=\epsilon/p_0-(\epsilon/p_0)^2\tfrac13\alpha_2,\quad
B:=\epsilon/q_0-(\epsilon/q_0)^2\tfrac13\alpha_2.
\end{equation}
The result is again the same as in \eqref{eq_A2-dbl}.

\subsection{C3}
We consider next the continuum limits of \eqref{eq:C34-x-only},
centered at $(0,0)$:
 { \begin{flalign*}
     &\frac{(x_{1,1}-x_{0,1})(x_{-1,1}-x_{0,0})(x_{-1,0}-x_{-1,-1})}
  {(x_{1,1}-x_{1,0})(x_{0,0}-x_{1,-1})(x_{0,-1}-x_{-1,-1})}=\nonumber&&\\
  &\frac{(x_{0,0}-x_{-1,1})( b_0\,x_{-1,0}+b_1)+(x_{0,1}\!-\!x_{-1,1})(x_{-1,0}\!-\!x_{0,0}) P
\!-\! (x_{0,1}\!-x_{0,0})(x_{-1,0}\!-x_{-1,1}) Q}
{(x_{1,-1}-x_{0,0})
    (b_0\,x_{0,-1}+b_1) +(x_{1,0}\!-\!x_{0,0})(x_{0,-1}\!-\!x_{1,-1}) P
    \!-\! (x_{1,0}\!-x_{1,-1})(x_{0,-1}\!-x_{0,0})Q}.&&
\end{flalign*}}

\subsubsection{Straight limit}
We use the usual straight limit approach \eqref{eq:str-def},
\eqref{eq:str-par}, and \eqref{eq:PQpq} but without any gauge
transformation. The leading term in the expansion is then
\begin{align}
&\partial_x\log\left[P(p)(u_1-u_{-1})
-b_1-b_0 u_{-1}+\frac{u_1'u_0'u_{-1}'}{(u_1-u_0)(u_0-u_{-1})}\right]\nonumber\\
&\hskip 3cm =\frac{u_1'}{u_1-u_0}-\frac{u_{-1}'}{u_0-u_{-1}}.\label{eq:C3-str}
\end{align}
A special case of this appears  in \cite{W-PhD-2001} (5.9.15).

\subsubsection{Skew limit}
For the skew limit we use \eqref{eq:skw-def}, \eqref{eq:skw-par} and
\eqref{eq:PQpq}, without gauge transform, and the leading term in the
expansion yields
\begin{align}
&\partial_\tau\log\left[P(p)(u_1-u_{-1}) -b_1-b_0
    u_{-1}%\Bigr.\nonumber\\& \hskip 1cm\Bigl.
    +\frac{(3p^2+2\alpha_2
    p+\alpha_1) (u_1-u_0)(u_0-u_{-1})}{\dot u_0}\right]\nn
    \\&\hskip5cm=\frac{\dot u_{-1}}{u_{-1}-u_{-2}}-\frac{\dot
      u_{1}}{u_2-u_{1}}.\label{eq:C3-skw}
\end{align}
This has some similarity to
the straight limit. When $P(p)=p^3,\,b_1=b_0=0$, one will
get (5.9.9) in \cite{W-PhD-2001} (up to signs).

\subsubsection{Double limit}
Starting from \eqref{eq:C3-str} and using $p=1/\delta$ and
$u_\nu=v(y+\nu\delta,x)$ with the additional variable change defined by
\[
\partial_y=(1-\tfrac13\alpha_2\delta)\partial_x+\tfrac12\delta\partial_t,
\]
we obtain
\begin{equation}\label{eq:C3-dbl}
3\partial_t\left(\frac{v_t}{v_x}\right)+\partial_x
\left(
\frac{v_{xxx}}{v_x}+\frac32\frac{v_t^2-v_{xx}^2}{v_x^2}
-4\frac{b_1+b_0 v}{v_x}
\right)=0.
\end{equation}

The double limit obtained by starting from the skew semi-continuous limit
\eqref{eq:C3-skw}, together with variable change
\[
\partial_y=(-1+\tfrac13\alpha_2\delta)\delta^2\partial_x+
\tfrac12\delta^3\partial_t,
\]
again yields \eqref{eq:C3-dbl}.

The direct double limit from \eqref{eq:C34-x-only} using
\eqref{eq:dbl-limi} with \eqref{eq:dbl-limi-A2}, but with the second
variable in \eqref{eq:dbl-limi} given by
$y-(n+m)\epsilon^2/(2p_0q_0)$, leads again to \eqref{eq:C3-dbl},
except for $b_1\mapsto -b_1$.

If $b_0=b_1=0$ in \eqref{eq:C3-dbl} it is the Schwarzian BSQ (see,
e.g., \cite{Weiss} Eq.~(4.9)), however the additional $b_0,b_1$ terms
break the M\"obius invariance.

\section{Lax pairs\label{S:lax}}
In general Lax pairs can be generated from CAC: One takes the side
equations and interprets the bar-shifted variables as linear Lax
variables. However, the equations defined on an edge only are not
convenient, thus one usually takes some linear combinations of them.
The construction of Lax matrices using CAC can be automatized to some
extent, see \cite{BHQK-FCM-2013,Bri,BHbk2}.

One important requirement for the Lax matrices is that they should
contain a spectral parameter, in the following it will be $R$.

\subsection{B2}
The B2 equations were given \eqref{B2-H}. On the left side of the
cube we then have
\begin{align}
  \t y=x\t x-z, \quad & \b y=x\b x-z,\\ \b{\t y}=\b x\b{\t x}-\b z,
  \quad & \b{\t y}=\t x\b{\t x}-\t z,\label{B2-yth}
\end{align}
and if from the second set we eliminate $\b{\t y}$ or $\b{\t x}$ we
get
\begin{equation}\label{eq:xompB3}
  \b{\wt{x}} = \frac{\wt{z}-\b{z}}{\wt{x}-\b{x}}\  ,\qquad
  \b{\wt{y}}=\frac{\wt{z}\,\b x-\b{z}\,\wt x}{\wt{x}-\b{x}}.
\end{equation}
Now we introduce \cite{ZZN-SAM-2012}
\begin{equation}\label{eq:laxphi}
\b x=\frac{\phi_1}{\phi_0}, \quad
\b z=\frac{\phi_2}{\phi_0}, \quad
\b y=\frac{\phi_3}{\phi_0},
\end{equation}
and then the equations in \eqref{eq:xompB3} can be written as
\bse
\begin{eqnarray}
\t \phi_0&=&\gamma(\t x \phi_0-\phi_1),\\
\t \phi_1&=&\gamma(\t z \phi_0-\phi_2),\\
\t \phi_3&=&\gamma(\t z \phi_1-\t x\phi_2),
\end{eqnarray}
where $\gamma$ is the separation factor. For the $\b{\wt{z}}$ equation
we take tilde-bar version of \eqref{B2-H-c} (with also $Q\to R$) and
eliminate from it $\b{\wt{x}}$ using \eqref{eq:xompB3}, this results
with
\begin{eqnarray}
\t \phi_2&=&\gamma(A_p\,\phi_0+(b_0\,x-b_1+y)\phi_1-(b_0+x)\phi_2),\\
\text{where}\quad A_p&:=&(b_1-b_0 x-y)\t x+(b_0+x)\t z+P-R.
\end{eqnarray}
\ese

The above can also be done for the hat-bar version. We then get the
matrix equations
\begin{equation}\label{eq:BSQ-LP}
\wt{\phi} =\bL\phi  \ ,
~~\wh{\phi} =\bM\phi  \ ,
\end{equation}
in which \cite{ZZN-SAM-2012}
\begin{equation}
\phi=\left(\begin{array}{c}\phi_0\\ \phi_1 \\ \phi_2 \\
\phi_3\end{array}\right),~~
\bL_{4\times 4}^{{B2}}=\gamma\left(
\begin{array}{cccc}
\wt{x} & -1 & 0 & 0\\
\wt{z} &  0 & -1 & 0\\
A_p & y+b_0 x-b_1 & -b_0 -x & 0 \\
0 &  \wt{z} & -\wt{x} & 0\\
\end{array}\right).
\label{eq:BSQ-L21}
\end{equation}
The matrix $\bM$ is the hat-$q$ version of $\bL$.  Now the
last column of $\bL,\bM$ is a null column and therefore we can reduce
the system to a $3\times 3$ matrix problem with
\begin{equation}
\phi=\left(\begin{array}{c}\phi_0\\ \phi_1 \\ \phi_2\end{array}\right),~~
\bL_{3\times 3}^{{B2}}
=\gamma\left(
\begin{array}{cccc}
\wt{x} & -1 & 0\\
\wt{z} &  0 & -1\\
A_p & y+b_0 x-b_1 & -b_0 -x
\end{array}\right).
\label{eq:BSQ-L213x3}
\end{equation}
The compatibility condition following from \eqref{eq:BSQ-LP} is
\begin{equation}\label{eq:compat}
  \h \bL \bM=\t\bM \bL.
\end{equation}
For this equation it would be best to choose the separation factor
$\gamma$ so that $\det \bL=$const., because then part of the matrix
equation is immediately satisfied by taking determinants of both
sides. In this case we can take $\gamma=1$ because $\det
\bL=P-R$.  With this choice \eqref{eq:compat} yields three equations:
\bse\begin{eqnarray} \h{\wt{x}} &=&
\frac{\wt{z}-\h{z}}{\wt{x}-\h{x}},\quad x=\frac{\t y-\h y}{\t x-\h
  x},\\ \h{\wt{z}} &=&
b_0(\h{\wt{x}}-x)+x\h{\wt{x}}+b_1-y+\frac{P-Q}{\t x-\h x}.
\end{eqnarray}\ese
This is not completely equivalent to \eqref{B2-H} because we do not
have un-shifted $z$, nor $\t z$ and $\h z$ separately.

In the language of DLA version of B2 \eqref{B2-ZZN} (with $b_1=0$),
the $3\times3$ Lax matrices are as follows,
\begin{equation}\label{eq:BSQ-Lk}
\bL^{B2D}=\left(
\begin{array}{ccc}
p-\wt u^{(0,0)} & 1 & 0\\
-\wt{u}{}^{(1,0)} & p-b_0 & 1 \\
\ast & -u^{(0,1)}-2b_0u^{(0,0)}-2b_0^2& p+2 b_0+u^{(0,0)}
\end{array}\right),
\end{equation}
in which
$\ast=R-P-(p-\wt{u}{}^{(0,0)})[(p+b_0)(p+u^{(0,0)})+u^{(0,1)}]-(p+u^{(0,0)}+2b_0)\wt{u}{}^{(1,0)}$,
and where $\bM^{B2D}$ is obtained from \eqref{eq:BSQ-Lk} by replacing
$p$ by $q$ and $\wt{\phantom{a}}$ by $\wh{\phantom{a}}$, furthermore
$\det [\bL^{B2D}]=R-P$.  The compatibility leads to the equations
\bse\be
\wh{u}{}^{(1,0)}-\wt{u}{}^{(1,0)}=(p-q+\wh{u}{}^{(0,0)}-\wt{u}{}^{(0,0)})
\wh{\wt{u}}{}^{(0,0)}-p\wh{u}{}^{(0,0)}+q\wt{u}{}^{(0,0)}\
\label{eq:u012-h-t_b} \ee
and \be\label{eq:u01}
\wh{u}{}^{(0,1)}-\wt{u}{}^{(0,1)}=(p-q+\wh{u}{}^{(0,0)}-\wt{u}{}^{(0,0)})u{}^{(0,0)}
-p\wt{u}{}^{(0,0)}+q\wh{u}{}^{(0,0)},
\ee\ese together with an equation that is equivalent to equation
\eqref{B2-ZZN-c}, if we use $\wh{\wt{u}}{}^{(0,0)}$ as solved from the
first equation and $u^{(0,0)}$ solved from the second.

The B2 Lax pair was first given in \cite{NPCQ-IP-1992} with
$\alpha_1=\alpha_2=b_0=b_1=0$ and in \cite{ZZN-SAM-2012} with full
parameterization.

\subsection{A2}
\paragraph{From CAC:}
The A2 equation was given in \eqref{A2-H}, where we can take
$b_0=0$. Using \eqref{A2-H-a} and the bar-version of \eqref{A2-H-b}
and their shifts, we can easily derive
\begin{equation}\label{eq:xompA3}
    \b{\wt{x}} = \frac{\wt{x}-\b{x}}{\wt{z}-\b{z}}\  ,\qquad
  \b{\wt{y}}=\frac{\wt{x}\,\b z-\b{x}\,\wt z}{\wt{z}-\b{z}}.
\end{equation}
To these and to the tilde-bar version of \eqref{A2-H-c} we use
\eqref{eq:laxphi} and obtain the Lax matrix \cite{ZZN-SAM-2012}
\begin{equation}
\bL=\gamma(x,\t x)\left(
\begin{array}{cccc}
 \wt{z} & 0 & -1& 0 \\ \wt{x} & -1 & 0 &
 0\\ \frac{y\wt{z}}{x}-\frac{P\wt{x}}{x} & \frac{R}{x}&
 -\frac{y}{x} & 0 \\ 0 & -\wt{z} & \wt{x}& 0
\end{array}\right).
\end{equation}
Again the last column vanishes and we use instead an invertible
$3\times 3$ matrix \cite{ZZN-SAM-2012}
\begin{equation}\label{Lax-a2-3}
\bL^{A2}=\gamma(x,\t x)\left(
\begin{array}{ccc}
 \wt{z} & 0 & -1\\ \wt{x} & -1 & 0
 \\ \tfrac{y\wt{z}}{x}+\tfrac{P\wt{x}}{x} & -\frac{R}{x} &
 -\tfrac{y}{x} \\
\end{array}\right),
\end{equation}
and the matrix $\bM^{A2}$ is the hat-$q$ version of $\bL^{A2}$. If we
again normalize the Lax matrix by the condition $\det [\bL]=$const,
we should take $\gamma(a,b)=(a/b)^{1/3}$. Then the compatibility
condition \eqref{eq:compat} yields
\begin{equation}\label{eq:C3compat}
\wh{\wt x}=\frac{\wt x-\wh x}{\wt z-\wh z}\  , \qquad x=\frac{\wh
x\wt y-\wt x\wh y}{\wt x-\wh x}\  ,
\end{equation}
together with eq. \eqref{A2-H-c}.

\paragraph{From DLA:}
In the DLA the A2 equation is given in \eqref{A2-ZZN} and the Lax
matrix given there is \cite{ZZN-SAM-2012}
\begin{equation}\label{eq:A2-L} \bL^{A2}=
\left( \begin{array}{ccccc} p-a && \wt{v}_a && 0 \\ 0 && p-\wt{u}{}^{(0,0)} &&
  1 \\ \frac{G_3(R,-a)}{v_a} && \ast && p-\alpha_2+\frac{s_a}{v_a}
\end{array}\right),
\end{equation}
in which $\ast=(p-\wt{u}{}^{(0,0)})(p-\alpha_2+s_a/v_a)-p_a\wt{v}_a/v_a$,
$R$ stands for the spectral parameter, and
$\bM^{A2}$ is obtained from \eqref{eq:A2-L} by replacing $p$ by $q$
and $\wt{\phantom{a}}$ by $\wh{\phantom{a}}$. (Recall that $v_a$ is
related to $x$, $u^{(0,0)}$ to $z$ and $s_a$ to $y$.) The Lax compatibility
condition leads to the equations \eqref{A2-ZZNc} and
\bse\label{eq:svurels} \begin{eqnarray}
  && p-q+\wh{u}{}^{(0,0)}-\wt{u}{}^{(0,0)}
  =(p-a)\frac{\wh{v}_a}{\wh{\wt{v}}_a}-(q-a)\frac{\wt{v}_a}{\wh{\wt{v}}_a}\ , \label{eq:svurels_b} \\
  \label{eq:svvrel}
&&p-q+\frac{\wh{s}_a}{\wh{v}_a}-\frac{\wt{s}_a}{\wt{v}_a}
=(p-a)\frac{v_a}{\wt{v}_a}-(q-a)\frac{v_a}{\wh{v}_a}\ .
\end{eqnarray}
\ese

\paragraph{From the two-component form:}
The A2 equation has also a two-component form,
\begin{subequations}
\label{A2-H-2c-No}
\begin{align}
&\frac{z-\t w\,}{z-\h w}=\frac{\h z-\th w\,}{\t z-\th w},
 \label{A2-H-2c-a-No}\\
&(\,\t z-\h z\,)\,(\,\th z-w)=\frac{P}{z-\t w}-\frac{Q}{z-\h w}. \label{A2-H-2c-b-No}
\end{align}
\end{subequations}
This pair is still defined on the elementary square and is
3-dimensionally consistent, but it is not possible to construct a Lax
pair using the sides of the consistency cube, because it leads to an
expression that is quadratic in the auxiliary field $\phi$. However, a
$2\times2$ Lax matrix was given in \cite{NongThesis} (see
eqn. (2.4.5)) (although this does not contain spectral parameter):
\begin{subequations}\label{A2-2c-LP}
\begin{align}
  \boldsymbol{L}_{2\times 2}^{A2}
&= \left(
     \begin{array}{cc}
         \t{z}  & -1\\
         w\t{z}+\frac{P}{z-\t w} & -w \\
     \end{array}
  \right),\\
  \boldsymbol{M}_{2\times 2}^{A2}
&= \left(
     \begin{array}{cc}
         \h{z}  & -1\\
         w\h{z}+\frac{Q}{z-\h w} & -w \\
     \end{array}
  \right),
\end{align}
\end{subequations}
which directly yields \eqref{A2-H-2c-No} from the compatibility $\h \bL
\bM=\t\bM \bL$.

Still another Lax matrix generating \eqref{A2-H-2c-No} is given in
\cite{FX-13} (see page 16)
\begin{equation}
\bL^{FX}=\left(
\begin{array}{ccc}
 \t{z}-w & 0 & -1\\
-1 &  \t w-z & 0 \\
\frac{P}{\t w -z}-(\t w-w)(\t z-w) & R & \t w-w \\
\end{array}\right),
\end{equation}
with corresponding $\bM^{FX}$. As mentioned above this Lax pair cannot
arise from CAC analysis of the type that worked for the other
equations. However, since this Lax matrix resembles some of the other
CAC-generated Lax matrices, we could try to reverse engineer and see
where it could come from. If we write out the $3\times3$ version of
$\wt{\phi}=\bL^{FX} \phi$ and then divide the second and third equation
by the first we get
\bse\label{eq:reveng}
\begin{eqnarray}
 \frac{{\wt\phi_1}}{\wt{\phi}_0}&=&
  \frac{1+(\t z-w)\frac{\phi_1}{\phi_0}}{w-\t z +\frac{\phi_2}{\phi_0}},\\
 \frac{ \wt{\phi}_2}{\wt{\phi}_0}&=&
  \frac{R\frac{\phi_1}{\phi_0}+(\wt{w}-w)\frac{\phi_1}{\phi_0}
+\frac{P}{\wt{w}-z}-(\wt{w}-w)(\wt{z}-w)  }{w-\t z +\frac{\phi_2}{\phi_0}}.
\end{eqnarray}
\ese
Previously we associated bar-quantities to $\phi$ ratios as in
\eqref{eq:laxphi} but it does not work here. We must instead take
\begin{equation}
  \frac{\phi_1}{\phi_0}=\frac1{\b w-z},\quad
  \frac{\phi_2}{\phi_0}=\b z-w,
\end{equation}
and this choice yields \eqref{A2-H-2c-No} from \eqref{eq:reveng}.

\subsection{C3}
The C3 equation \eqref{C3-H} is
\bse\label{eq:xyz-MSBSQ-1-3DC}
\begin{eqnarray}
  && x-\wt{x} = \wt{y}z\  , ~~ x-\wh{x} = \wh{y}z\  ,   \label{eq:xyz-MSBSQ-1-3DCa} \\
&& \wh{\wt{z}}\,y = z\frac{P\wh{z}\wt{y}-Q\wt{z}\wh{y}}
     {\wt{z}-\wh{z}}+b_1+b_0\,x\ .  \label{eq:xyz-MSBSQ-1-3DCb1}
\end{eqnarray}\ese
Several other equations can be derived from
\eqref{eq:xyz-MSBSQ-1-3DCa}, for example
    \begin{equation} \label{eq:altC3}
\wh{\wt{x}} = \frac{\wh{x}\wt{z}-\wt{x}\wh{z}}{\wt{z}-\wh{z}}\ ,~~x
= \frac{\wh{x}\wt{y}-\wt{x}\wh{y}}{\wt{y}-\wh{y}} \ , ~~
z = -\frac{\wt{x}-\wh{x}}{\wt{y}-\wh{y}}\ ,\quad
\wh{\wt{y}}
=-z\frac{\t y-\h y}{\t z-\h z}.
    \end{equation}

If we consider the tilde-bar versions of these and take equation for
$\b{\wt{x}}$ from \eqref{eq:altC3} and equation
\eqref{eq:xyz-MSBSQ-1-3DCb1} in which $\t y$ and $\b y$ have been
eliminated using \eqref{eq:xyz-MSBSQ-1-3DCa}, then we can use the CAC
method and construct the $3\times 3$ Lax matrix
\begin{equation}\label{Lax-C3}
\bL_{3\times3}^{C3}= \frac1{z}\begin{pmatrix} \t z&0&-1\\ 0&\t z&-\t
  x\\ \tfrac{\t z}{y}(b_1+(b_0-R)\,x)& \tfrac{\t z}{y}R
  &\tfrac1{y}(P(\t x-x)-b_1-b_0\,x)
\end{pmatrix}.
\end{equation}
The compatibility conditions arising from this Lax matrix and its
hat-Q companion yield the equations for $\h{\t x}$ and $x$ in
\eqref{eq:altC3}. However, for $\h{\t z}$ it produces an equation
that agrees with \eqref{eq:xyz-MSBSQ-1-3DCb1} only after we also use
the equation for $z$ in \eqref{eq:altC3}, which does not follow from
Lax compatibility.

There is also the $4\times 4$ Lax matrix that can be obtained from CAC by adding the $\wh{\wt{y}}$ equation from \eqref{eq:altC3}:
\begin{equation}
\bL_{4\times4}^{C3}= \frac{1}{z}\left(
\begin{array}{cccc}
\wt{z} & 0 & 0 & -1 \\
0 & \wt{z} & 0 & -\wt{x} \\
-z\wt{y} &  0 & z & 0 \\
\frac{\wt{z}}{y}(b_1+b_0\,x)  & 0 & -\frac{z\wt{z}}{y}R
& \frac{1}{y} (P z\wt{y}-b_1-b_0\,x)\\
\end{array}\right),
\end{equation}
and the matrix $\bM_{4\times 4}^{C3}$ is the hat-$Q$ version. In this
case the compatibility conditions yield equation
\eqref{eq:xyz-MSBSQ-1-3DCb1} and equations for $\h{\t x},\,x,\,\h{\t y}$ in
\eqref{eq:altC3}.

If $b_0=0$ we also have the two-component version
(\ref{C3-H-b1-c},\ref{C3-H-b1-d}) containing $z$ and $y$ only. Using
the CAC method we can construct the $3\times 3$ Lax matrix
\begin{equation}\label{Lax-C3b1}
\bL_{3\times3}^{C3b_1}=
\frac1{z}\begin{pmatrix}
  \t z&0&-1\\
  -z\t y&z&0\\
b_1\frac{\wt{z}}{y} & -R\frac{z\wt{z}}{y}& \frac1y(Pz\t y-b_1)
\end{pmatrix}
\end{equation}
and with corresponding $\bM_{3\times3}^{C3b_1}$ the compatibility
condition exactly returns (\ref{C3-H-b1-c}, \ref{C3-H-b1-d}).

\section{Bilinear structures of DBSQ-type equations}\label{sec-4}

\subsection{Preliminary }\label{sec-2}

\subsubsection{Discrete Hirota's bilinear form and Casoratians}

Suppose $f_j(n,m)$ and $g_j(n,m)$ are functions defined on
$\mathbb{Z}\times \mathbb{Z}$.  Then a one-component discrete Hirota
bilinear equation has the following
form \cite{HZ-JPA-2009,HZ-JDEA-2013}
\begin{equation*} \label{eq:HB} \sum_j\,
c_j\, f_{j}(n+\nu_{j}^+,m+\mu_{j}^+)\,
g_{j}(n+\nu_{j}^-,m+\mu_{j}^-)=0,
\end{equation*}
where it is essential that the index sums $\mu_{j}^++\mu_{j}^-=\mu_s$,
$\nu_{j}^++\nu_{j}^-=\nu_s$ do not depend on $j$.

The solutions to a discrete Hirota bilinear equation will be given by a
{\bf Casoratian}, which is a determinant of a matrix composed of different
shifts of a vector. For example, given functions $\psi_i(n,m,l)$ we
define the column vector
\begin{equation}\label{C-entry-vec}
\psi(n,m,l)=\bigl(\psi_1(n,m,l),\psi_2(n,m,l),\cdots,\psi_{N}(n,m,l)\bigr)^T,
\end{equation}
and then the $N$-th order Casoratian reads
\begin{equation}
  \label{eq:C-gen}
    \bigl|\psi(n,m,l_1),\psi(n,m,l_2),\cdots,\psi(n,m,l_N)\bigr|.
\end{equation}
For such a determinant we use a shorthand notation $|l_1,l_2,\cdots,
l_N|$.  Furthermore, if the Casoratian contains consecutive columns we
use condensed notation such as (cf. \cite{FN-PLA-1983-KP}),
$$|0,1,\cdots,N-1| =  |\h{N-1}|,~~|0,1,\cdots,N-2,N| = |\h{N-2},N|.$$

For the DBSQ-type equations discussed in this paper the solutions can
be expressed through a Casoratian $f= |\h{N-1}|$ which is composed by
the {\em entry function} $\Psi$:
\begin{eqnarray}
\psi_j(l,n,m,\alpha,\beta)&:=&\sum^{2}_{s=0} (-\omega_s (k_j))^l
(p-\omega_s (k_j))^n (q-\omega_s (k_j))^m\nn \\ &&\hskip 2.2cm
\times (a-\omega_s
(k_j))^\alpha(b-\omega_s (k_j))^\beta\,\varrho^{(0)}_{j,s},
\label{psi-gen-0}
\end{eqnarray}
which contains 5 independent variables $n,m,l,\alpha$ and $\beta$.  In
the following we do not mention those variables that are obvious from
context.  Here $\omega_s (k_j),\ s=1,2$ are roots of
\begin{equation}
g_3(\omega(k_j))=g_3(k_j),
\end{equation}
where $g_3$ defined in \eqref{ga} and $\omega_0 (k_j) \equiv k_j $.

For shifts in the index variables $(l,n,m,\alpha,\beta)$ we often use
shorthand notation: in addition to tilde and hat for shifts in the
$n$- and $m$-direction (as in \eqref{th}), we introduce bar, circle and
dot for the shifts in the $l$-, $\alpha$- and $\beta$-directions,
respectively, i.e.
\bse\label{shiftdef}\begin{align}
&\b f(n,m,l,\alpha,\beta)=f(n,m,l+1,\alpha,\beta),\\
&\c f(n,m,l,\alpha,\beta)=f(n,m,l,\alpha+1,\beta),\\
&\d f(n,m,l,\alpha,\beta)=f(n,m,l,\alpha,\beta+1).
\end{align}
\ese
When the symbol is below the variable it means backward shift: e.g.,
$\ut{f}(n,m,l,\alpha,\beta) = f(n-1,m,l,\alpha,\beta)$.

A more general form than \eqref{psi-gen-0} is
\begin{equation}
  \psi_j(l)=\sum^{2}_{s=0} (\delta-\omega_s (k_j))^l (p-\omega_s
  (k_j))^n (q-\omega_s (k_j))^m (a-\omega_s (k_j))^\alpha(b-\omega_s
  (k_j))^\beta\,\varrho^{(0)}_{j,s},
\label{psi-gen-del}
\end{equation}
which is referred to as the $\delta$-extension of \eqref{psi-gen-0}.
Thus \eqref{psi-gen-del} can be considered as a function containing
symmetrically five dimensions, corresponding to the direction
coordinates $(n,m,l,\alpha,\beta)$ and their lattice spacing
parameters $(p,q,a,b,\delta)$.  Here for $\psi_j(l)$
in \eqref{psi-gen-del} we already omitted $n,m,\alpha,\beta$ for
convenience, we will often do this if it does not cause any confusion.

Finally we note that a Casoratian can sometimes be written as a
Wronskian. Suppose we define \begin{equation}
\phi_j(l)=\sum^{2}_{s=0}
  e^{(\delta-\omega_s (k_j))l} (p-\omega_s (k_j))^n (q-\omega_s (k_j))^m
  (a-\omega_s (k_j))^\alpha(b-\omega_s (k_j))^\beta\,\varrho^{(0)}_{j,s},
\label{phi-gen-del}
\mathcal{}\end{equation}
and the corresponding vector as in \eqref{C-entry-vec}.
Then the following Casoratian and Wronskian are equal to each other,
\[
|\psi(l_1),\,\psi(l_2),\cdots,\psi(l_N)|_{\mathrm{C}[\psi]}
=|\partial_l^{l_1}\phi(l),\,\partial_l^{l_2}\phi(l),\cdots,
\partial_l^{l_N}\phi(l)|_{\mathrm{W}[\phi]}.
\]

\subsubsection{Laplace expansion}
The proof that a Casoratian solves a bilinear equation is usually
given by reducing the problem to a three-term Laplace expansion of a
zero determinant.  We will first give a generic result.

\begin{lemma}\label{L:lap}
Suppose that $\mathbf{P}$ is a $N\times(N-1)$ matrix, and $\mathbf{Q}$
its $N\times(N-k+1)$ sub-matrix obtained by removing arbitrary $(k-2)$
columns from $\mathbf{P}$ where $k \geq 3$. Let $\mathbf{a}_i (i=1,
2, \cdots, k)$ be some $N$-th order column vectors.  Then we have
\begin{equation}
\sum^{k}_{i=1}(-1)^{i-1} |\mathbf{P}, \mathbf{a}_i| \,
     |\mathbf{Q}, \mathbf{a}_1, \cdots, \mathbf{a}_{i-1},
     \mathbf{a}_{i+1}, \cdots, \mathbf{a}_{k}|=0.
\label{lap}
\end{equation}
\end{lemma}
This is a special case of Pl\"uker relations.

In fact, let $\mathbf{B}$ be the $N \times (k-2)$ matrix consisting of
those $(k-2)$ column vectors that are removed from $\mathbf{P}$ so
that $|\mathbf{QB}|=|\mathbf P|$.  Then it is easy to see that the
following $2N \times 2N$ determinant vanishes:
\begin{equation}
\left|
  \begin{array}{cccccc} \mathbf{Q} & \mathbf{0} & \mathbf{B}
    & \mathbf{a}_1 & \cdots & \mathbf{a}_k \\ \mathbf{0} & \mathbf{Q}
    & \mathbf{0} & \mathbf{a}_1 & \cdots & \mathbf{a}_k \\ \end{array}
\right| = 0.
\label{zero-det}
\end{equation}
The LHS of \eqref{lap} is actually the Laplace expansion of the
LHS of \eqref{zero-det}.

Equation \eqref{lap} is quite useful in proving solutions. When
$k=3$, \eqref{lap} yields
\begin{equation}
 |\mathbf{P},\mathbf{a}_1|\,|\mathbf{Q},\mathbf{a}_2,\mathbf{a}_3|
-|\mathbf{P},\mathbf{a}_2|\,|\mathbf{Q},\mathbf{a}_1,\mathbf{a}_3|
+|\mathbf{P},\mathbf{a}_3|\,|\mathbf{Q},\mathbf{a}_1,\mathbf{a}_2|=0.
\label{lap-k3-1}
\end{equation}
If we denote $\mathbf{P}=(\mathbf{Q},\mathbf{a}_0)$, it reads
\begin{equation}
 |\mathbf{Q},\mathbf{a}_0,\mathbf{a}_1|\,|\mathbf{Q},\mathbf{a}_2,\mathbf{a}_3|
-|\mathbf{Q},\mathbf{a}_0,\mathbf{a}_2|\,|\mathbf{Q},\mathbf{a}_1,\mathbf{a}_3|
+|\mathbf{Q},\mathbf{a}_0,\mathbf{a}_3|\,|\mathbf{Q},\mathbf{a}_1,\mathbf{a}_2|=0,
\label{lap-k3-2}
\end{equation}
which can be viewed as an expression of Pl\"uker relation, and were
used to verify Wronskian (or Casoratian) solutions to bilinear
equations \cite{FN-PLA-1983-KP} and also to prove many determinantal
identities (see \cite{Gragg-SIAM-rev-1972}).  When $k=4$, \eqref{lap}
yields
\begin{align}
&|\mathbf{P},\mathbf{a_1}|
   \,|\mathbf{Q},\mathbf{a_2},\mathbf{a_3},\mathbf{a_4}|
 -|\mathbf{P},\mathbf{a_2}|
   \,|\mathbf{Q},\mathbf{a_1},\mathbf{a_3},\mathbf{a_4}|
 +|\mathbf{P},\mathbf{a_3}|
   \,|\mathbf{Q},\mathbf{a_1},\mathbf{a_2},\mathbf{a_4}| \nonumber\\
 -& |\mathbf{P},\mathbf{a_4}|
   \,|\mathbf{Q},\mathbf{a_1},\mathbf{a_2},\mathbf{a_3}|
 =0,
\label{lap-k4}
\end{align}
which is also useful in solution verification
(see \cite{ZZZ-PLA-2009}).  In this paper, besides \eqref{lap-k3-1}
and \eqref{lap-k4}, we need also $k=5$ case, which is
\begin{align}
 &|\mathbf{P},\mathbf{a_1}|
   \,|\mathbf{Q},\mathbf{a_2},\mathbf{a_3},\mathbf{a_4},\mathbf{a_5}|
 -|\mathbf{P},\mathbf{a_2}|
   \,|\mathbf{Q},\mathbf{a_1},\mathbf{a_3},\mathbf{a_4},\mathbf{a_5}|
 +|\mathbf{P},\mathbf{a_3}|
   \,|\mathbf{Q},\mathbf{a_1},\mathbf{a_2},\mathbf{a_4},\mathbf{a_5}|\nonumber\\
 -&|\mathbf{P},\mathbf{a_4}|
   \,|\mathbf{Q},\mathbf{a_1},\mathbf{a_2},\mathbf{a_3},\mathbf{a_5}|
 +|\mathbf{P},\mathbf{a_5}|
   \,|\mathbf{Q},\mathbf{a_1},\mathbf{a_2},\mathbf{a_3},\mathbf{a_4}|
  =0.
\label{lap-3}
\end{align}

Here is another determinantal property which is often used in
Wronskian/Casoratian verification.
\begin{lemma}\cite{FN-PLA-1983-KP}
\label{L:iden}
\begin{equation}\label{iden-1}
\sum_{j=1}^{N}|\mathbf{a}_1,\cdots,\mathbf{a}_{j-1},\,
\mathbf{b}\mathbf{a}_{j},\, \mathbf{a}_{j+1},\cdots,
\mathbf{a}_{N}|=\biggl(\sum_{j=1}^{N}b_{j}\biggr)|\mathbf{a}_1,\cdots,
 \mathbf{a}_N|,
\end{equation}
where ~$\mathbf{a}_j=(a_{1j},\cdots,a_{Nj})^T$ and
$\mathbf{b}=(b_1,\cdots,b_N)^T$ are ~$N$-th order column vectors and
$\mathbf{b}\mathbf{a}_j$ stands for ~$(b_{1}a_{1j}, \cdots, b_N
a_{Nj})^{T}$.
\end{lemma}
A generalized version of this lemma can be found
in \cite{ZDJ-arxiv-2006,ZZSZ-RMP-2014}.
For further details see  \cite{Zhang-Wr-2019}.

\subsection{B2}\label{sec-4.1}
\subsubsection{B2}
Equation B2 (\eqref{B2-H} with $b_1=0$) is given by
\begin{subequations}\label{DB2}
\begin{align}
B_1:=\quad  & \t y - x\t x+z=0,
  \label{DB-a}\\
B_2:=\quad  & \h y - x\h{x}+z=0,
  \label{DB-b}\\
B_3:=\quad & y - b_0(\th x-x)-x\th{x}+\th{z}-\frac{P-Q}{\t{x}-\h{x}}=0,\label{DB-c}
\end{align}
where
\begin{equation}
P=p^3-b_0 p^2+R,\quad Q=q^3-b_0 q^2+R.
\label{abb0-pq}
\end{equation}
\end{subequations}
It has background solution \cite{HZ-JMP-2010}
\begin{subequations}
\begin{align}
x_0 &= pn+qm+c_1,\label{x0}\\
z_0 &= \tfrac{1}{2}x^2_0+\tfrac{1}{2}(p^2n+q^2m+c_2)+c_3,\label{z0}\\
y_0 &=\tfrac{1}{2}x^2_0-\tfrac{1}{2}(p^2n+q^2m+c_2)-c_3,\label{y0}
\end{align}
\label{0SS-HB-B2}
\end{subequations}
where $c_1,c_2,c_3$ are arbitrary constants.

By the dependent variable transformation \cite{HZ-JMP-2010}
\begin{equation}
x=x_0-\frac{g}{f},~~~z=z_0-x_0 \frac{g}{f}+\frac{h}{f},~~~
y=y_0-x_0 \frac{g}{f}+\frac{s}{f},
\label{trans}
\end{equation}
we can bilinearize the B2 lattice consisting of \eqref{DB2}
and their shifts follows
\begin{equation}
  B_1 =\frac{\mathcal{B}_1}{f\t{f}}\,,
  ~ B_2  =\frac{\mathcal{B}_2}{f\h{f}}\,,~
  B_3
  =\frac{\mathcal{B}_3\mathcal{B}_4+(p-q)f\th{f}\mathcal{B}_4
+[p^2+pq+q^2-b_0(p+q)]\t{f}\h{f}\mathcal{B}_3}{(\t x-\h x)f\t{f}\h{f}\th{f}},
\end{equation}
where the bilinear equations are
\begin{subequations}
\label{bil-HDB-I}
\begin{align}
  \mathcal{B}_1&:=\t{f}(h+pg)-\t{g}(g+pf)+f\t{s}=0, \label{bil-HDB-a}\\
  \mathcal{B}_2&:=\h{f}(h+qg)-\h{g}(g+qf)+f\h{s}=0,\label{bil-HDB-b}\\
  \mathcal{B}_3&:=\t{f}\h{g}-\h{f}\t{g}+(p-q)(\t{f}\h{f}-f\th{f})=0,
\label{bil-HDB-c}\\
\mathcal{B}_4&:=[p^2 \!+\!pq\! +\!q^2\!-\!b_0(p\!+\!q)](f\th{f}\!-\!\t{f}\h{f})\!+\!
(p\!+\!q\!-\!b_0)(\th f g\!-\!f \th g)\!+\!\th{f}s\!+\!f\th{h}\!-\!g\th{g}\!=0.\label{bil-HDB-d}
\end{align}
\end{subequations}
We have also used the parametrization \eqref{abb0-pq}.

The set of bilinear equations \eqref{bil-HDB-I} admits $N$-soliton
solutions in the following Casoratian form,
\begin{equation}
\label{casdef}
  f=|\h{N-1}|,\,  g=|\h{N-2},N|,
\,  h=|\h{N-2},N+1|,\, s=|\h{N-3},N-1,N|,
\end{equation}
composed of $\psi=(\psi_1,\psi_2,\cdots,\psi_N)^T$
with \eqref{psi-gen-0}.  A proof for these Casoratian solutions can be
found in \cite{HZ-SIGMA-2011}, where the meaning of the parameter $b_0$
is also discussed.

An alternate bilinearization was given in \cite{MK}, where they first
transformed the 9-point equation \eqref{eq:B2-x-only} without $b_0$
into a pair of equations living on a $2\times 4$ stencil (Eqs.~(53,
54)), which were then bilinearized using only $f$ and $g$
(see Eqs.~(56, 57, 65, 66)).

\subsubsection{B2-$\delta$}\label{b2del}
In the above derivation we used Casoratians with
entries \eqref{psi-gen-0}.  However, there is a natural generalization
of the entry function with a new parameter $\delta$ as given
in \eqref{psi-gen-del}. We will now derive the corresponding
generalized equations.

As the first step we compute the difference of $f,g,h,s$ (as defined
in \eqref{casdef}) for $\delta=0$ and for $\delta\neq0$.  Using
binomial expansion on $(\delta-\omega_s (k_j))^l$ we see that only
the right-most columns contribute and find \cite{HZ-JMP-2010}
(here $f\equiv f(0),\,f'\equiv f(\delta))$
\bse\label{fghsdel}\begin{eqnarray}
f'&=&f ,\\ g'&=&g +N \delta\, f ,\\
h'&=&h +(N+1)\delta\, g +\tfrac12 N(N+1)\,\delta^2f ,\\
s'&=&s +(N-1)\delta\, g +\tfrac12 N(N-1)\,\delta^2f ,
\end{eqnarray}
\ese
where $N$ is the size of the determinant. Now inverting this for
$f,g,h,s$ and using them on \eqref{bil-HDB-I} it follows that the
primed quantities solve the $\delta$-modified bilinear equations
\begin{subequations}
\label{bil-B2-II}
\begin{align}
  \mathcal{B}^{\delta}_1
&:=\t{f'}\,[\,h'+(p-\delta)g'\,]-\t{g'}\,[\,g'+(p-\delta)f'\,]+f'\,\t{s'}=0,
\label{bil-B2-a2}\\
  \mathcal{B}^{\delta}_2
&:=\h{f'}\,[\,h'+(q-\delta)g'\,]-\h{g'}\,[\,g'+(q-\delta)f'\,]+f'\,\h{s'}=0,
\label{bil-B2-b2}\\
  \mathcal{B}^{\delta}_3
&:=\t{f'}\,\h{g'}-\h{f'}\,\t{g'}+(p-q)\left(\t{f'}\,\h{f'}-f'\th{f'}\,\right)=0,
  \label{bil-B2-c2}\\
  \mathcal{B}^{\delta}_4
&:=(p^2+pq+q^2-b_0(p+q))\left(f'\th{f'}-\t{f'}\,\h{f'}\,\right)\nonumber\\&\quad
 -(p+q-b_0+\delta)\left(f'\,\th{g'}-\th{f'}g'\right)
 +\th{f'}\,s'+f'\,\th{h'}-g'\,\th{g'}=0.\label{bil-B2-d2}
\end{align}
\end{subequations}
Furthermore, from \eqref{trans} we get
\bse
\begin{equation}
x'=x'_0-\frac{g'}{f'}=(x'_0-N\delta)-\frac{g}{f}=x,
\end{equation}
provided that $x'_0=x_0+N\delta$, which can be accommodated by a change
in the constant: $c'_1=c_1+N\delta$.  Similarly
\begin{eqnarray}
z'&=&z'_0-x'_0 \frac{g'}{f'}+\frac{h'}{f'}\nn\\
&=&z'_0-(x_0+N\delta) \frac{g}{f}-x'_0
N\delta+ \frac{h}{f}+(N+1)\delta\frac{g}{f} +\tfrac12
N(N+1)\,\delta^2\nn\\
&=&z-\delta(x-x_0),\\
y'&=&y+\delta(x-x_0),
\end{eqnarray}
\ese
provided that the constants in $z_0,\,y_0$ are adjusted so that
$\tfrac12(c_2'-c_2)+c_3'-c_3+\tfrac12\delta^2N=0$.
Using these we get the $\delta$-modified nonlinear equations:
\begin{subequations}
\label{B2-II}
\begin{align}
B_1^\delta &:= \t {y'}- x'\t {x'}+z'-\delta\,\left(\,\t{x'}-x'-p\right)=0,\label{B2-II-a}\\
B_2^\delta &:= \h {y'}- x'\h{x'}+z'-\delta\,\left(\,\h{x'}-x'-q\right)=0,\label{B2-II-b}\\
B_3^\delta &:=
y'\!-\!b_0\! \left(\th{x'}\!-\!x'\right)\!-\! x'\th{x'}+\th{z'}\!
+\frac{p^3\!-\!q^3\!-\!b_0(p^2\!-\!q^2)}{\h{x'}-\t{x'}}
+\!\delta\! \left(\th{x'}\!-\!x'\!-\! p-\! q\! \right)\!=0.
\label{B2-II-c}
\end{align}
\end{subequations}
Equations \eqref{bil-B2-II} and \eqref{B2-II} were given for the
$b_0=0$ case in \cite{HZ-JMP-2010} and for generic $b_0$
in \cite{NongZSZ2013}.

\subsection{A2}\label{sec-4.2}

\subsubsection{A2}
The A2 equation (after removing parameter $b_0$) is given by
\begin{subequations}
\label{DB-A2-a}
\begin{align}
A_1:=\quad  &\t x z -\t y -x=0,\label{A2-a1}\\
A_2:=\quad  &\h x z -\h y -x=0,\label{A2-b1}\\
A_3:=\quad  &x \th z -y-
\frac{P\t{x}-Q\h{x}}{\t{z}-\h{z}}=0,\label{A2-c1}
\end{align}
\end{subequations}
where $P,\,Q$ will be parameterized by \eqref{A2-PQ-DLA}.  It has
background or seed solution \eqref{eq:xyz-0-A2}
\begin{subequations}\label{A2-xyz0}
\begin{align}
x_a &= (p-a)^{-n} (q-a)^{-m} c_1,\label{x-0}\\
z_0 &= (c_3-p)n+(c_3-q)m+c_2,\label{z-0}\\
y_a &=x_a(z_0+a-c_3),\label{y-0}
\end{align}
\label{0SS-HB-A2}
\end{subequations}
where $c_3=\alpha_2/3$, and  $c_1,c_2,\alpha_i$ are arbitrary constants.

By the dependent variable transformation
\begin{equation}\label{A2-trans}
x=x_a\frac{\dc f}{f},~~~z=z_0+ \frac{g}{f},~~~ y=y_a
\frac{\dc f}{f}+x_a\frac{\dc g}{f},
\end{equation}
A2 is bilinearized into
\begin{subequations}
\label{bil-A2-a}
\begin{align}
\mathcal {A}_1&:= \tdc{f}(g+pf)-(p-a)\dc{f}\t{f}-f(a\tdc{f}+\tdc{g})=0,\label{bil-A2-a-1}\\
\mathcal {A}_2&:= \hdc{f}(g+qf)-(q-a)\dc{f}\h{f}-f(a\hdc{f}+\hdc{g})=0,\label{bil-A2-a-2}\\
\mathcal {A}_3&:= \t{f}\h{g}-\h{f}\t{g}\,+\,(p-q)(\t{f}\h{f}-f\th{f})=0,\label{bil-A2-a-3}\\
\mathcal {A}_4&:=
(p-q)[((p+q+a-\alpha_2)\dc{f}+\dc{g})\th{f}-\th{g}\dc{f}]
-p_a\h{f}\tdc{f}+q_a\t{f}\hdc{f}=0,\label{bil-A2-a-4}
\end{align}
\end{subequations}
where $p_a,\,q_a$ are defined as \eqref{paqa}
 and the circle shift was defined in \eqref{shiftdef}.
The connection to \eqref{DB-A2-a} is by
\begin{equation}\label{A2-bil-rela}
  A_1 =\frac{\t x_a\mathcal{A}_1}{f\t{f}},
~
  A_2 =\frac{\h x_a\mathcal{A}_2}{f\h{f}},
~
  A_3 =x_a\frac{\mathcal{A}_3\mathcal{A}_4+(p-q)f\th{f}\mathcal{A}_4
               +(p_a\h{f}\tdc{f}-q_a\t{f}\hdc{f})\mathcal{A}_3}
              {(p-q)f\t{f}\h{f}\th{f}(\h z-\t z)}.
\end{equation}

Multisoliton solutions to \eqref{bil-A2-a} are given through
\begin{equation}
\label{casdef-A2}
f=|\h{N-1}|,\,  g=|\h{N-2},N|,
\end{equation}
composed of $\psi$ given in \eqref{psi-gen-del} with $\delta=0$.

\subsubsection{A2-$\delta$}
The $\delta$ deformation discussed in Section \ref{b2del} can also be
applied on A2. In this case the definitions of $f,g$ are as in
\eqref{casdef}, and therefore the $\delta$-dependence is as in
\eqref{fghsdel}, i.e., $f'=f ,\, g'=g +N \delta\, f .$ From this and
\eqref{A2-trans} it follows
\begin{equation}\label{eq-trans-A2}
x=x', \quad y=y', \quad z=z'-\delta,
\end{equation}
provided that we choose $c_2'-c_2+N\delta=0$. Then we find
A2-$\delta$ as
\begin{subequations}\label{DB-A2-II}
\begin{align}
A_1^\delta &:= \t {x'} z' -\t {y'} -x'-\delta \t {x'}=0,\label{A2-a2}\\
A_2^\delta &:= \h {x'} z' -\h {y'} -x'-\delta \h {x'}=0,\label{A2-b2}\\
A_3^\delta &:= x' \th {z'}-y'-\delta x'
   +\frac{P\t{x'}-Q\h{x'}}{\t{z'}-\h{z'}}=0.
\label{A2-c2}
\end{align}
\end{subequations}
On the other hand, it can be easily seen that equations
\eqref{bil-A2-a} are invariant under $g\mapsto g+c\,f$.

\subsection{C3$_{b_0}$}\label{sec-4.3}
The equation C3$_{b_0}$ is
\begin{subequations}\label{C3-b0}
\begin{align}
C_1 & := \t x - \t y \,z - x=0 ,\label{C3-a}\\
C_2 & := \h x - \h y \,z - x=0 ,\label{C3-b}\\
C_3 & := \th{z}\,y - d_2\,x
        - z \frac{P\,\t y\,\h z-Q\,\h y\,\t z}{\t z-\h z}=0.\label{C3-c}
\end{align}
\end{subequations}
It has 0SS
\begin{subequations}\label{C3-0ss}
\begin{align}
 & x_{a,b}= \frac{1}{b-a}\left(\frac{p-b}{p-a}\right)^n
                        \left(\frac{p-b}{q-a}\right)^m,\label{C3-x0}\\
& z_b= (p-b)^{n}(q-b)^{m},\label{C3-z0}\\
& y_a= -(p-a)^{-n}(q-a)^{-m},\label{C3-y0}
\end{align}
\end{subequations}
where we have used parametrization \eqref{C3-PQb0}.
By the transformation (c.f., \eqref{shiftdef})
\begin{equation}\label{C3-trans}
x=x_{a,b}\frac{\d{\dc f}}{f}, \quad
z=z_b \frac{\d f}{f}, \quad
y=y_a \frac{\dc f}{f},
\end{equation}
from \eqref{C3-b0} we have its bilinear form
\begin{subequations}
\label{bil-C3}
\begin{align}
\mathcal {C}_1&:= (p-b)\tddc{f} f-(p-a)\ddc{f} \t{f}+(b-a)\d{f} \tdc{f}=0,
\label{bil-C3-a}\\
\mathcal {C}_2&:= (q-b)\hddc{f} f-(q-a)\ddc{f} \h{f}+(b-a)\d{f} \hdc{f}=0,
\label{bil-C3-b}\\
\mathcal {C}_3&:= (p-b)\h{f}\td{f}-(q-b)\t{f}\hd{f}+(p-q)\th{f}\d{f}=0,\label{bil-C3-c}\\
\mathcal {C}_4&:=
(p-q)\thd{f}\dc{f}+\frac{(p-q)a_b}{(p-b)(q-b)}\th f \ddc{f}
-\frac{p_a}{p-b}\hd{f}\tdc{f}+\frac{q_a}{q-b}\td{f}\hdc{f}=0,\label{bil-C3-d}
\end{align}
\end{subequations}
where $p_a,\,q_a$ are defined as \eqref{paqa}
and here we also use $a_b=b_a=p_a|_{p=b}$,
and the connection with \eqref{C3-b0} is
\begin{equation}\label{C3-bil-rela}
\begin{split}
  &C_1 =\frac{-x_{a,b}\mathcal{C}_1}{(p-a)f\t{f}}\,,
\quad
   C_2 =\frac{-x_{a,b}\mathcal{C}_2}{(q-a)f\h{f}}\,,
\\
  &C_3 =\frac{(p-b)(q-b)(b-a)x_{a,b}z_b}
             {(p-q)f\t{f}\h{f}\th{f}(\t z-\h z)}\!
        \left[\mathcal{C}_3\mathcal{C}_4
            \! +\!\left(
               \frac{p_a}{p-b}\hd{f}\tdc{f}
              -\frac{q_a}{q-b}\td{f}\hdc{f}
              \right)\!\mathcal{C}_3
            \! -(p-q)\th f\d{f}\mathcal{C}_4\! \right] .
\end{split}
\end{equation}

Casoratian solution of \eqref{bil-C3} is given by $f=|\h{N-1}|$ with
$\psi$ composed by \eqref{psi-gen-del}.  Note that C3-$\delta$ is the
same as \eqref{C3-b0}.

The bilinearization of C3$_{b_1}$ is still open.

\section{Conclusions}\label{sec-5}
In this review we have discussed the fully discrete versions
Boussinesq equations given as three-component equations on the basic
quadrilateral. Their derivation using CAC and DLA was compared. From
the three-component equation we derived two- and one-component
versions on a larger stencil. Then we derived two semi-continuous
limits and the fully continuous limits for the one-component versions.
We discussed also several versions of Lax pairs. Finally we gave their
Hirota bilinear forms, which are important for constructing
solutions. The basic results are summarized in the adjoining Table
\ref{T1} which points to the relevant equations.

We also note that recently an elliptic scheme of DLA has been developed,
in which the spacing parameters $P$ and $Q$ can be parameterized by
Weierstrass elliptic functions \cite{NSZ-elliptic}.

In addition to the question of bilinearizing equation C3$_{b_1}$,
there are some interesting open questions: For example, the classification
of higher order CAC lattice equations (cf. the ABS list containing 9
equations), higher genus solutions of the DBSQ-type equations,
reductions of the hierarchy of lattice KP equations, etc.

\setlength{\fboxsep}{1pt}
\renewcommand{\arraystretch}{1.4}

\begin{table}
\begin{center}
  \begin{tabular}{|l|c|c|c|c|}
    \hline Equation name & A2 & B2 & C3$_{b_0}$ & C3$_{b_1}$\\ \hline
    3 component versions & \fbox{\eqref{A2-H}}, \fbox{\eqref{A2-Hw}}&
    \fbox{\eqref{B2-H}} & \fbox{\eqref{C3-H-1}} & $\leftarrow$
    \\ \hline 2 component versions &
    \eqref{eq:A2-xy},\fbox{\eqref{eq:A2-yz}} & \eqref{eq:B2-xy}
    \fbox{\eqref{RSB2}} & \eqref{eq:Cxy}, \fbox{\eqref{eq:C3-2-yz}} &
    \fbox{\eqref{eq:Csimple-both}}\\ \hline 1 component versions &
    \eqref{eq:A2-x-only}, \eqref{eq:A2-z-only} & \eqref{eq:B2-x-only}
    &\eqref{eq:C34-x-only}, \eqref{eq:A2-x-only} & $\leftarrow$
    \\ \hline Straight limit & \eqref{eq:A2-str} & \eqref{B2-CL-st1} &
    \eqref{eq:C3-str} & $\leftarrow$\\ \hline Skew limit &
    \eqref{eq:A2-skw} & \eqref{eq:B2skew} & \eqref{eq:C3-skw} &
    $\leftarrow$\\ \hline Double limit &\eqref{eq_A2-dbl} &
    \eqref{eq:B2-dbl} & \eqref{eq:C3-dbl} & $\leftarrow$\\ \hline Lax
    Pairs & \eqref{Lax-a2-3}+others & \eqref{eq:BSQ-L213x3},
    \eqref{eq:BSQ-Lk} & \eqref{Lax-C3} & \eqref{Lax-C3b1}\\\hline
    Bilinear form & \eqref{bil-A2-a} & \eqref{bil-HDB-I} &
    \eqref{bil-C3} & ?\\ \hline
  \end{tabular}
\end{center}
  \caption{A summary of the relevant equations.  Those defining
    equations that live on the elementary quadrilateral are boxed. The
    ``$\leftarrow$'' in the last column means the result is included in
    the C3$_{b_0}$ case.\label{T1}}
\end{table}

\section*{Acknowledgments}
 We would like to thank J. Schiff for bringing equation
  \eqref{RSB2} to our attention.  This work was supported by the NSF
  of China (grant numbers 11875040 and 11631007). All computations were
  done with REDUCE \cite{Hearn}.

\begin{appendices}

  \section{Triply shifted variables from CAC}
  Since the equations are integrable their triply shifted forms are
  the same independent of the sides used in the computation, therefore
  tilde-hat-bar symmetric. They can be computed if the equations are
  defined on a quadrilateral. This holds for all three-component forms
  and for some two-component forms but the one component forms are all
  defined on a bigger stencil.

  \subsection{Three-component forms}
Three-component forms all live in the elementary quadrilateral and
were in fact derived using CAC. The triply shifted quantities are as
follows \cite{H-JPA-2011}:

For A2
\begin{subequations}
\begin{eqnarray}
\b{\h{\t x}}&=&x\,\frac{
\t x(\h z-\b z)+
\h x(\b z-\t z)+
\b x(\t z-\h z)}{
p\t x(\h z-\b z)+
q\h x(\b z-\t z)+
r\b x(\t z-\h z)},\\
\b{\h{\t y}}&=&\frac{y}x\, \b{\h{\t x}}+ \frac{
p\t x(\h x-\b x)+
q\h x(\b x-\t x)+
r\b x(\t x-\h x)}{
p\t x(\h z-\b z)+
q\h x(\b z-\t z)+
r\b x(\t z-\h z)},\\
\b{\h{\t z}}&=&z-x\frac{
p(\h z-\b z)+
q(\b z-\t z)+
r(\t z-\h z)}{
p\t x(\h z-\b z)+
q\h x(\b z-\t z)+
r\b x(\t z-\h z)}.
\end{eqnarray}
\end{subequations}

For B2
\begin{subequations}
\begin{eqnarray}
\b{\h{\t x}}&=&b_0+x+\frac{
(q-r)\t x+
(r-p)\h x+
(p-q)\b x}{
\t x(\h z-\b z)+
\h x(\b z-\t z)+
\b x(\t z-\h z)},\\
\b{\h{\t y}}&=&b_0x+y+\frac{
(q-r)\t z+
(r-p)\h z+
(p-q)\b z}{
\t x(\h z-\b z)+
\h x(\b z-\t z)+
\b x(\t z-\h z)},\\
\b{\h{\t z}}&=&z+b_0\b{\h{\t x}}-\frac{
(q-r)\h x\b x+
(r-p)\b x\t x+
(p-q)\t x\h x}{
\t x(\h z-\b z)+
\h x(\b z-\t z)+
\b x(\t z-\h z)}.
\end{eqnarray}
\end{subequations}

For C3
\begin{subequations}
\begin{eqnarray}
\b{\h{\t x}}&=&\frac{b_0x+b_1}{y}\b{\h{\t y}}+x+z\frac
{\t z\h y\b y(q-r)+\h z\b y\t y(r-p)+\b z\t y\h y(p-q)}
{\t z(q\h y-r\b y)+\h z(r\b y-p\t y)+\b z(p\t y-q\h y)},\\
\b{\h{\t y}}&=&y\frac{
\t z(\h y-\b y)+\h z(\b y-\t y)+\b z(\t y-\b y)}
{\t z(q\h y-r\b y)+\h z(r\b y-p\t y)+\b z(p\t y-q\h y)},\\
\b{\h{\t z}}&=&b_0z+(b_1+b_0 x)\frac{(q-r)\t z+(r-p)\h z+(p-q)\b z}
{\t z(q\h y-r\b y)+\h z(r\b y-p\t y)+\b z(p\t y-q\h y)\nn}\\
&&+zy\frac{
qr\t z(\h y-\b y)+rp\h z(\b y-\t y)+pq\b z(\t y-\b y)}
{\t z(q\h y-r\b y)+\h z(r\b y-p\t y)+\b z(p\t y-q\h y)}.
\end{eqnarray}
\end{subequations}

\subsection{Two-component forms}
When the two component form is defined on a quadrilateral we can
compute the triply shifted quantities. This is true for
\eqref{eq:A2-yz},\eqref{eq:Csimple-both},\eqref{eq:C3-2-yz}.

For A2 \eqref{eq:A2-yz} where $w:=y/x$ \bse
\begin{align} \xbht w
=& w +\frac{P(\b w-\h w)+Q(\t w-\b w)+R(\h w-\t w)} {\t z(\t w-z)(\h
    w-\b w)+\h z(\h w-z)(\b w-\t w)+ \b z(\b w-z)(\t w-\h w) },\\
  \xbht z =&\\
  &\frac{P\t w(\h w\!-\!z)(\b w\!-\!z)(\h z\!-\!\b z)
\!+\!  Q\h w(\b w\!-\!z)(\t w\!-\!z)(\b z\!-\!\t z)\!
+\!  R\b w(\t w\!-\!z)(\h w\!-\!z)(\t z\!-\!\h z)}
  {P(\h w-z)(\b w-z)(\h z-\b z)
\!+\!  Q(\b w\!-\!z)(\t w-z)(\b z-\t z)\!
+\!  R(\t w-z)(\h w-z)(\t z-\h z)}.\nn
\end{align}
\ese

For B2 \eqref{RSB2} where $w:=y+z$
\bse\begin{align} \xbht x=& b_0 +
\frac{P(\b x-\h x)+Q(\t x-\b x)+R(\h x-\t x)}
     {\t w(\b x-\h x)+\h w(\t x-\b x)+\b w(\h x-\t x)},\\
     \xbht w=&\xbht x(b_0-x)+w+x^2+\nn\\
     &\frac{
P(\b w-\h w+(\b x-\h x)\t x)
\!+\!Q(\t w-\b w+(\t x-\b x)\h x)
\!+\!R(\h w-\t w+(\h x-\t x)\b x)
}{\t w(\b x-\h x)+\h w(\t x-\b x)+\b w(\h x-\t x)}.
\end{align}
\ese

For C$3_{b_1}$ \eqref{eq:Csimple-both}:
\bse\begin{align} \xbht y =&
  -y\frac{\t y(\h z-\b z)+\h y(\b z-\t z)+\b y(\t z-\h z)}
  {P\t y(\h z-\b z)+Q\h y(\b z-\t z)+R\b y(\t z-\h z)},\\
 \xbht z =&
  -z\frac{PQ\b z(\t y-\h y)+RP\h z(\b y-\t y)+QR\t z(\h y-\b y)}
  {P\t y(\h z-\b z)+Q\h y(\b z-\t z)+R\b y(\t z-\h z)}\nn\\
&  +b_1\frac{P(\h z-\b z)+Q(\b z-\t z)+R(\t z-\h z)}
{P\t y(\h z-\b z)+Q\h y(\b z-\t z)+R\b y(\t z-\h z)}.
\end{align}
\ese

For  C$3_{b_0}$ \eqref{eq:C3-2-yz}:
\bse\begin{align}
\xbht w=-w&\frac{
 (\t w\h w z+\b w)(\t z -\h z)
+(\h w\b w z+\t w)(\h z -\b z)
+(\b w\t w z+\h w)(\b z -\t z)}{\Cal D_w},\\
      {\Cal D_w}=&
  \t w(P+b_0)(z(\h w\h z-\b w\b z)-\h z+\b z)
+ \h w(Q+b_0)(z(\b w\b z-\t w\t z)-\b z+\t z)\nn\\
&+\b w(R+b_0)(z(\t w\t z-\h w\h z)-\t z+\h z),\\
\xbht z=\phantom{-i}z&\frac{{\Cal N_z}}{{\Cal D_z}},\\
      {\Cal N_z}=&
  PQ(\t w-\h w)(\b w z\!-\!1)\b z
+ QR(\h w-\b w)(\t w z\!-\!1)\t z
+ RP(\b w-\t w)(\h w z\!-\!1)\h z\nn\\
&+ b_0[
 P(\h wz-1)(\b wz-1)(\h z-\b z)
+Q(\b wz-1)(\t wz-1)(\b z-\t z)\nn\\&\phantom{+b_0(}
+R(\t wz-1)(\h wz-1)(\t z-\h z)],\\
{\Cal D_z}=
 &\phantom{b_0[}P(\h wz-1)(\b wz-1)(\h z-\b z)\t w
+Q(\b wz-1)(\t wz-1)(\b z-\t z)\h w\nn\\&\phantom{+b_0(}
+R(\t wz-1)(\h wz-1)(\t z-\h z)\b w.
\end{align}
\ese

Note that none of these results have the tetrahedron property.

\end{appendices}

\strut\hfill

%\noindent
%    {\bf A Remark on the references:}
%    The references should be ordered alphabetically according to the
%authors in the list and the authors' initials should be after their surname without a
%comma, e.g. Ovsienko V and Roger C. Please see below.

\end{document}